----------------------------- Start of body part 2

\documentstyle[12pt]{report}

\begin{document}
\pagestyle{empty}
\begin{titlepage}

\title{Spin-Anisotropy Commensurable Chains:
Quantum Group Symmetries and N = 2 SUSY }
\author { Alexander B\'erkovich, C\'esar G\'omez and Germ\'an Sierra \\
\\
Instituto de Matem\'aticas y F\'{\i}sica Fundamental, CSIC \\
Serrano 123.  Madrid, SPAIN \\ \\
January 1993}
\date{}
\maketitle

\begin{abstract}

In this paper we consider a class of the 2D integrable models. These models
are higher spin $XXZ$ chains with an extra condition of the commensurability
between spin and anisotropy. The mathematics underlying this commensurability
is provided by the quantum groups with deformation parameter being an $N$th
root of unity. Our discussion covers a range of topics including new integrable
deformations, thermodynamics, conformal behaviour, $S$-matrices and
magnetization. The emerging picture strongly depends on the $N$-parity. For
the $N$ even case at the commensurable point, $S$-matrices factorize into
$N=2$ supersymmetric Sine-Gordon matrix and an RSOS piece.
The physics of an $N$ odd case is rather different. Here, the
supersymmetry does not manifest itself and the bootstrap hypothesis fails.
Away from the commensurable point, we find an unusual magnetic behaviour. The
magnetization of our chains depends on the sign of the external magnetic field.

\end{abstract}
\end{titlepage}
\pagestyle{plain}

\chapter*{Introduction and Discussion}

Symmetry is the driving concept in particle physics.
In Quantum Field Theory the particles are defined as finite dimensional
irreps of the space-time and internal symmetry groups. In Statistical
Mechanics the notion of symmetry has also played a very
important role in the past, as a way of characterizing degrees of freedom and
types of interaction. The recently introduced quantum groups are another
stride in this direction, which deepens our understanding of symmetry
in systems with an infinite number of degrees of freedom. Quantum groups
shed new light on the number of difficult problems which are, essentially,
non perturbative in nature; determination of the particle spectrum, scattering
matrices, correlation functions, to name just a few.

An important family of integrable models is based on the quantum
affine algebras $U_q(\hat{G})$ [1]. These models are characterized
by the quantum deformation parameter $q$ and a finite
dimensional irrep of $\hat{G}$. The famous six-vertex model
corresponds to $\hat{G} = \widehat{Sl(2)}$ and the fundamental spin
1/2 irrep. \cite{2}. The higher spin versions of
$U_q(\widehat{Sl(2)})$ \cite{3}, \cite{4} can be constructed by fusion
procedure \cite{5}.
Generally, the spectrum of these models is expected to satisfy
the bootstrap axioms \cite{6}: there is a fundamental particle
such that all others can be interpreted as bound states. For the
isotropic models of spin $j$ ($ q= 1$) the particle spectrum
consists of a fundamental spin 1/2 particle which has (for $j >1/2)$
extra hidden RSOS spin \cite{7}, \cite{8}. The corresponding physical
$S$-matrix factorizes into the product of the $XXX$ $S$-matrix
and an RSOS piece. This picture remains essentially unchanged
in the semiclassical weak anisotropic regime \cite{8}, \cite{9}.
The central extension for these nearly isotropic models is
$c = \frac{3 j}{j +1}$, i.e. that of a  $SU(2)_k$
$WZW$ model with level $ k = 2j $. Moreover, the excitations above the
ground state can be understood in terms of the free bosonic and
$Z_{2j}$-parafermionic \cite{10} degrees of freedom \cite{11}.

There is a connection between $N=2$ integrable field theories in two
dimensions and quantum affine algebras at roots of unity. This
became clear in the study of the Sine-Gordon model in ref. \cite{12}
and more recently in \cite{13}. The $N=2$ structure of the Sine-Gordon
models appears at the special value $ \beta^2= 8 \pi 3/2$
which yields $q^4 = 1$ ( $q= e^{i \pi /p}$ ,
$p = \beta^2/( 8 \pi - \beta^2)$) with the SUSY
realized in a non local way \cite{14}, namely, the one defined
by the non local action of the quantum group generators \cite{15}.
The solitonic $S$-matrix of the $N=2$ theories
factorizes into an $N=0$ RSOS $S$-matrix  and the $N=2$ $S$-matrix
of the Sine-Gordon model at $\beta^2= 8 \pi 3/2$ \cite{16}.
Observe that this factorization of the $S$-matrices is similar
to the one obtained for the physical $S$-matrices of higher spin $XXX$ models
using Bethe ansantz techniques. The only difference is that the $XXX$
piece of the $S$-matrix
is replaced in the case of the $N =2$ models by the Sine-Gordon
$XX$ $S$-matrix (i.e. the one which corresponds to $q^4 =1$).
One could wonder at this point whether there exist spin chains whose
$S$-matrices factorize as those of the $N=2$ models. We shall give
an example of these later.

Commensurability for $U_q(\widehat{Sl(2)})$-models appears
whenever the deformation parameter $q$ is an $N$th-root
($q^N =1$), and the dimension of the irrep used in constructing
the model is $N'$  ($N'= N$ ($N/2$) if $N$ odd (even)).
For $N$ even the commensurable point determines the frontier
of the weak anisotropic regime. The first result that we
obtain comes from the explicit solution of the Bethe equations
for the $N$ even case. From a technical point of view the novelty
of the Bethe ansatz analysis is related to the fact that $2j+1$ is not a
so called Takahashi number, an assumption that is
done in previous studies of higher spin models \cite{9}, \cite{17}, \cite{24}.
The elementary excitations $S$-matrix has the same structure as that of
$N=2$ SUSY models, namely, a Sine-Gordon $XX$ $S$-matrix multiplied by an
RSOS matrix associated with graph $A_{2j +2}$. This fact suggests that the
commensurable spin $j$-chain with $q=e^{ \frac{i \pi}{2j+1}}$ provides
a lattice model of $N=2$ integrable theories.

There is another important reason to study the spin-anisotropy commensurable
hamiltonians, namely, the peculiarities
of representation theory of quantum groups at root of unity \cite{18}.
At these values of $q$ there appear new central elements, that is,
new casimirs which form a closed Hopf subalgebra and whose
continuous spectrum labels the irreps. Thus, one can change in a
continuous way the irrep without changing its dimension. The
models so obtained are the non-fusion descendants of the six-vertex
model without classical $q =1$ analog. The Chiral Potts model
constructed from a class of irreps called cyclic, is one example of the
latter \cite{19}. In this paper we shall limit our attention
to a very special class of the non-classical irreps which we call nilpotent
\cite{20}, \cite{21}, \cite{21.5}. These irreps have the highest and lowest
weights. For the nilpotent deformations of the $N$ even commensurable model
we find a similar spectrum and the same central extension as for the
commensurable point. The main difference is a replacement of the $XX$ part
of the physical $S$-matrices by the nilpotent $R$-matrix with $N=4$. This
nilpotent $R$-matrix is related to an $XX$ chain placed in the external
magnetic field. However, the RSOS piece of the whole $S$-matrix remains
unchanged. The $N=4$ nilpotent $R$-matrix discussed above is, naturally,
associated with the quantum algebra $U_{\hat{q}}(Gl(1,1))$,
where $\hat{q}$ is fixed by the casimir which describes the nilpotent
irreps. Since the $Gl(1,1)$ algebra underlies $N=2$ SUSY, we suggest
that our nilpotent case is related to the $\hat{q}$-deformed $N=2$ SUSY.

Another issue of interest appears while studying the behaviour of
the nilpotent chains under the action of external magnetic field.
The external magnetic field acts on all degrees of freedom, while
the "nilpotent" magnetic field acts effectively only on the $XX$ part.
The net effect is that the magnetization depends on the sign
of the external magnetic field. The magnitude of this effect is proportional
to the ratio of the $XX$ degrees of freedom to the total number of degrees of
freedom (i.e. it decreases for higher $N$).

We extend our analysis to commensurable spin chains with
$q^N =1$ and $N$ odd. The picture we obtain is drastically different
from that of the $N$ even case. The particle spectrum consists of the three
types of particles whose energies do not satisfy bootstrap relations.
The scattering shifts among these particles are the product of an RSOS piece
associated with graph $A_{2 j_{eff}+2}$ with $j_{eff} =j/2$ (notice the
renormalization
of the spin) and another RSOS part associated with the "spin 1" model
$p=2,r=6$ \cite{22}. The (p=2,r=6) RSOS model captures most of the
aspects of our model: counting of states, central extensions
and scattering shifts. However, there is a mismatch in
energies, which reveals some failure of the bootstrap.
For the $N$ odd case we find no trace of the $N=2$ SUSY structure, but
this is not surprising, since SUSY always requires an even number
of degrees of freedom to be present. An attractive possibility will be to
interpret the $N$ odd case in terms of a $Z_3 $-parafermionic
extension of $N=2$ SUSY \cite{16}.

\noindent
Let us summarize the main results we obtained:
\noindent
i) Integrable deformations of the spin chains away from the commensurable
point. \\
ii) Extension of the Bethe ansantz analysis to the cases where $2s +1$
is not a Takahashi number. \\
iii) Computation of the $S$-matrices for the $N$ even and odd cases,
and connection with $N=2$ SUSY in the $N$ even case. \\
iv) New method of counting degrees of freedom in the vicinity
of the ground state. \\
v) Unusual Magnetic properties of the nilpotent models due to a non trivial
interplay between an "internal" magnetic field associated with the nilpotent
casimirs and the external magnetic field.

The organization of the paper is as follows. In section 1 we
present the basic tools for solving the descendent models
including a brief description of the algebraic Bethe ansantz.
Section 2 contains some basics on the representation theory of
quantum groups at roots of unity and a detailed description of the
symmetries, $R$-matrices and hamiltonians based on the nilpotent
irreps. In section 3 we discuss the string hypothesis and solve
the Bethe equations for the values of the nilpotent parameter,
which yield the hermitian hamiltonians. In section 4 we study the $N$ even
case, computing the central extensions and the $S$-matrices with
$N =2$ SUSY structure. In section 5 we extend this analysis to
the N odd case. Finally, in section 6 we study the magnetic
properties which depend on the direction of the external magnetic field.
The paper ends with concluding remarks and two appendices where technical
details are given.

\chapter*{1. Six-vertex Model Descendants}

\setcounter{chapter}{1}
\setcounter{section}{0}
\setcounter{equation}{0}
\section{Descendent Models}

Our starting point will be the quantum affine algebra
$U_q(\widehat{Sl(2)}) \equiv U_q(A^{(1)}_1)$ \cite{1}. The finite dimensional
irreps of this algebra at level zero can be easily obtained
from the finite dimensional irreps of the non affine quantum algebra
$U_q(Sl(2))$ whose generators E, F and K satisfy the well known relation:
\begin{eqnarray}
E\;K = q^2 K\;E\;;\;\;\; &   F\;K = q^{-2} K\;F\;;\;\;\;
&\left[ E, F \right] = \frac{K - K^{-1}}{q - q^{-1}}
\label{2.1}
\end{eqnarray}

Denoting by $\pi_{j}^{\pm}$ an irrep of $U_q(Sl(2))$ of spin $j$
of positive or negative spin parity we have:

\begin{eqnarray}
\pi_j^{\pm}(E) e_r & =& \left[r \right] e_{r-1}\nonumber \\
\pi_j^{\pm}(F) e_r& =& \pm \left[2j-r \right]  e_{r+1}
\label{2.2}\\
\pi_j^{\pm}(K) e_r&  =& \pm q^{2j-2r} e_r\nonumber
\end{eqnarray}

\noindent
where $r=0,1,\dots, 2j$.
This representation can be promoted to a finite dimensional irrep
$\pi_{j,u}^{\pm}$ of $U_q(\widehat{Sl(2)})$ as follows:
\begin{eqnarray}
\pi_{j,u}^{\pm}(E_0)= e^u \pi^{\pm}_j(F) &
\pi_{j,u}^{\pm}(E_1)= e^u \pi^{\pm}_j(E) & \nonumber \\
\pi_{j,u}^{\pm}(F_0)= e^{-u} \pi^{\pm}_j(E) &
\pi_{j,u}^{\pm}(F_1)= e^{-u} \pi_j^{\pm}(F) &
\label{2.3} \\
\pi_{j,u}^{\pm}(K_0)=  \pi_j^{\pm}(K^{-1}) &
\pi_{j,u}^{\pm}(K_1)=  \pi_j^{\pm}(K) & \nonumber
\end{eqnarray}

\noindent
where $u$ is called the spectral parameter.
Given two irreps $\pi_{j_1,u_1}$ and $\pi_{j_2,u_2}$ the intertwiner
matrix $R^{j_1,j_2}(u_1 -u_2)$ is defined by:

\begin{equation}
R^{j_1,j_2}(u_1-u_2) \; \pi_{j_1,u_1} \otimes \pi_{j_2,u_2}
\Delta(a) = \pi_{j_1,u_1} \otimes \pi_{j_2,u_2}
\Delta'(a) R^{j_1,j_2}(u_1 -u_2)
\label{2.4}
\end{equation}

The intertwiners (\ref{2.4}) satisfy the Yang-Baxter equation with spectral
parameter:

\begin{equation}
R_{12}^{j_1,j_2}(u_1-u_2) \;R_{13}^{j_1,j_3}(u_1-u_3) \;
R_{23}^{j_2,j_3}(u_2-u_3)=
R_{23}^{j_2,j_3}(u_2-u_3) \;R_{13}^{j_1,j_3}(u_1-u_3) \;
R_{23}^{j_1,j_2}(u_1-u_2)
\label{2.5}
\end{equation}

In fact, the intertwiner $R$-matrices, obtained through solving (\ref{2.4}),
are the trigonometric
solutions to the Yang-Baxter equation (\ref{2.5}) and give rise
to the integrable vertex models if one identifies the Boltzmann
weights with the intertwiners $R^{j_1,j_2}$
\footnote{ Writing $ u$ as $hu$ and $q$ as $e^h$ and taking the
limit $ h \rightarrow 0$ one gets the rational solutions depending on $ u$.}.
The simplest vertex model where the previous construction applies is nothing
but the six-vertex model \cite{2} which corresponds to the choice
$j_1=j_2=1/2 $.
Taking into account the fact that the spin 1/2 is a fundamental irrep of
$ U_q (Sl(2)) $, one calls the $(j_1,j_2)$-models descendants
of the six-vertex model. Given the intertwiner $R ^{j_1,j_2} (u)$ one
proceeds, in the spirit of \cite{23}, to construct the monodromy and transfer
matrices:

i)Monodromy matrix:

\begin{equation}
T^{j_1,j_2}(u) = R^{j_1,j_2}(u)_{a,L}
R^{j_1,j_2}(u)_{a,L-1}\cdots R^{j_1,j_2}(u)_{a,1}
\label{2.6}
\end{equation}

\noindent
where $L$ is the number of sites of the chain and $T^{j_1,j_2}$ is a
$(2j_1+1)\times (2j_1+1)$ matrix which belongs to $End( \otimes^L V^{j_2})$.

ii) Transfer matrix:
\begin{equation}
t^{j_1,j_2}(u) = Tr( T^{j_1,j_2}(u))
\label{2.7}
\end{equation}

{}From the Yang-Baxter equation (\ref{2.5}), which is a consequence of the
quasitriangularity of $U_q(\widehat{ Sl(2)})$, one gets the commutativity
of the transfer matrices:

\begin{equation}
\left[ t^{j_1,j}(u_1), t^{j_2,j}(u_2) \right] = 0
\label{2.8}
\end{equation}

\noindent
which implies, in particular, that the transfer matrices of the $(j,\;j)$-model
can be simultaneously diagonalized together with the one of the
$(1/2,\;j)$-model.
The main advantage of this result is that the algebraic Bethe ansantz can be
easily extended to the $(j,\;j)$-models.

\section{ Algebraic Bethe Ansantz for Higher Spin}

Let us, first, consider the case of a quantum deformation
parameter $ q$ not equal to the root of unity. We shall briefly summarize
the algebraic Bethe ansantz for a vertex model of spin $j$. The
basic tools that one needs are :

i) Reference state:

\begin{eqnarray}
& | \Omega_j> = e^j_0 \otimes \cdots \otimes e^j_0 & \nonumber\\
& e^j_0 =\left( \begin{array}{c}
1 \\ 0 \\ \vdots \\ 0
\end{array} \right) & \in  C^{2j+1}
\label{2.91}
\end{eqnarray}

ii) Spin-wave creation operator:

\begin{equation}
B_j (u) = \left[T^{1/2,j}(u) \right]_{1,0}
\label{2.10}
\end{equation}

Recall that $T^{1/2,j}(u)$ is a $2 \times 2$ matrix. The action of the
$B_j(u)$ operator on the reference state (\ref{2.91}) reduces the
total spin of this state by one and creates a spin wave whose quasimomenta
depends on the spectral parameter $u$.

iii) Spin-wave annihilation operator:

\begin{equation}
C_j(u) = \left[ T^{1/2,j}(u) \right]_{0,1}
\label{2.11}
\end{equation}

\noindent
Since the spin $j$ irrep is a highest weight vector irrep it follows:

\begin{equation}
C_j(u) \; |\Omega_j> = 0
\label{2.12}
\end{equation}

Thanks to (\ref{2.8}) the candidates for the eigenvectors of
the transfer matrix $t^{1/2,j}$ are of the form
$\prod^m_{i=1} B_j(u_i)$:

\begin{equation}
(A_j(u)+D_j(u)) \; \prod^m_{i=1}
B_j(u_i) |\Omega_j> = \Lambda_j (u,u_i) \prod_{i=1}^m B_j(u_i)
|\Omega_j>
\label{2.13}
\end{equation}

where
\begin{eqnarray}
A_j(u) = (T^{1/2,j}(u))_{0,0} &, & D_j(u) = (T^{1/2,j}(u))_{1,1}
\label{2.14}
\end{eqnarray}

The conditions on the spin-wave rapidities $u_i$ imposed by (\ref{2.13})
can be easily obtained using the commutation relations between
the creation operators $B_j(u)$ and $A_j(u)$ and $D_j(u)$. These
commutation relations are encoded in the famous RTT=TTR equation
which follows from the definition of the monodromy operators and
the Yang-Baxter relation (\ref{2.5}):

\begin{equation}
R_{12}^{j_1 j}\;  T_1^{j_1 j}\; T_2^{j_2 j}= T_2^{j_2 j} \;
T_1^{j_1 j}\; R_{12}^{j_1 j_2}
\label{2.15}
\end{equation}

\noindent
where $T_1=T\otimes 1$, $T_2=1\otimes T_2$. \\
Taking $j_1=j_3=1/2,\;j_2=j$, we get the desired
commutation relations. Indeed, equation (\ref{2.15}) defines a so called
Yang-Baxter algebra whose "structure constants" are determined
by the six-vertex $R$-matrix $R^{1/2,1/2}$.
Since the irrep $\pi_j$ is a h.w.v. representation, one deduces that
$A_j(u)$ and $D_j(u)$ are (2j+1) $\times$ (2j+1) diagonal matrices.
Defining $a_j(u)$ and $b_j(u)$ as:

\begin{eqnarray}
A_j(u) |\Omega_j> = (a_j(u))^L |\Omega_j> &, &
B_j(u) |\Omega_j> = (b_j(u))^L |\Omega_j>
\label{2.16}
\end{eqnarray}

\noindent
the Bethe equations for the rapidities $u_i$ read:

\begin{equation}
\left( \frac{ a_j (u_l) }{ b_j(u_l) } \right)^L =
\prod^M_{k \neq l, k=1} \frac{a(u_k-u_l) b(u_l-u_k)}{
a(u_l-u_k) b(u_k-u_l)}
\label{2.17}
\end{equation}

\noindent
where $a(u)=a_{1/2}(u)$ and $b(u)=b_{1/2}(u)$.
Notice that only the l.h.s. of equation (\ref{2.17}) depends on
$j$, the r.h.s., corresponding to the six-vertex model, is the
same for the whole family of descendants.
It will be convenient for later use to introduce the following change of
variables: $ u \rightarrow \frac{\gamma}{2} (u-i)$. In terms of these
new variables, the Bethe equations become:

\begin{equation}
\left( \frac{sinh \frac{\gamma}{2} (u_l + 2j\;i)}{
sinh \frac{\gamma}{2} (u_l - 2j\;i)} \right)^L=
\prod^M_{k=1,k \neq l}
\frac{sinh \frac{\gamma}{2} (u_l - u_k +2i)}{
sinh \frac{\gamma}{2} (u_l - u_k - 2i)}
\label{2.18}
\end{equation}

The hamiltonian of the spin chain is obtained through
the logarithm derivative of the transfer matrix:
\begin{equation}
H_{j}(\gamma) = iI\frac{\partial}{\partial u} log t^{j,j}(u)|_{u=0}
\label{2.19}
\end{equation}

\noindent
where $I$ is an overall coupling constant. \\
For $j=1/2$ one gets the well known $XXZ$-chain with anisotropy
$\Delta = \frac{ q+q^{-1}}{2} = cos\gamma$. In the limit, when
the parameter $\gamma$ goes to zero, one recovers from
(\ref{2.19}) the higher spin isotropic models of the first two
refs. in \cite{4}. An example of the spin 1 isotropic hamiltonian is:

\begin{equation}
H_{j=1}(\gamma =0) = \sum_{i=1}^L S_i \cdot S_{i+1} - ( S_i S_{i+1})^2
\label{2.20}
\end{equation}

Within the family of anisotropic higher spin integrable models
we shall pay particular attention to the
cases where the anisotropy $\gamma$ and the spin $j$ are commensurable
in the sense:

\begin{eqnarray}
q^N =1, & 2j +1 =N, & N\; odd \nonumber \\
q^N =1, & 2j +1 = N/2, & N\; even
\label{2.21}
\end{eqnarray}

\noindent
where $q=e^{i\gamma}$.
At these points, the higher spin $XXZ$ models admit continuous integrable
deformations parametrized by the eigenvalues of the central
elements of $U_{q}(\widehat{Sl(2)})$ at $q^N=1$. This will be the
subject of the next section.

\chapter*{2. New integrable Deformations for $q^N=1$}

\setcounter{chapter}{2}
\setcounter{section}{0}
\setcounter{equation}{0}

\section{Mathematical Preliminaries: Representation Theory of
$U_q(Sl(2))$}

The representation theory of quantum groups at roots of unity
has been worked out in full generality in refs. \cite{17.5}, \cite{18}.
Here, we
will only need the results concerning $U_q(Sl(2))$, which are summarized below.

For $q^N =1$ the Hopf algebra $U_q(Sl(2))$ contains a central
Hopf subalgebra ${\cal{Z}}_q$ generated by the elements $E^{N'}
,F^{N'}$ and $K^{N'}$, where $N'= N$ ($N/2$) if $N$ is odd (even).
The Schur lemma implies that the finite dimensional irreps
$\pi$ are characterized by the eigenvalues of the central elements:

\begin{equation}
\pi(a) = \xi_{\pi} (a) 1 \;\; ,\;\;  a \in {\cal{Z}}_q
\label{3.1}
\end{equation}

The finite dimensional irreps can then be divided, according
to the eigenvalues of the central elements, into the following
four types:

\begin{center}
\begin{tabular}{|c|c|c|c|c|c|}
\hline
irrep $\pi$ &$ \xi_{\pi}(E^{N'})$ &$ \xi_{\pi}(F^{N'})$
 &$ \xi_{\pi}(K^{N'})$
& dimension & hwv / lwv \\
\hline
cyclic  &$ x \neq 0$ &$ y \neq 0$ &$ z \neq \pm 1$ &$N'$& no/no  \\
\hline
semicyclic & 0  & y$ \neq 0$ &$ z \neq \pm 1$ &$N'$& yes/no  \\
\hline
nilpotent & 0 & 0 & generic  &$N'$ & yes/yes \\
\hline
classical & 0 & 0 &$ \pm 1$ &$ \leq N' $ & yes/yes  \\
\hline
\end{tabular}
\begin{center}
Table 1.
\end{center}
\end{center}

The last two columns reflect the highest and the lowest weight
properties of the irreps. The classical irreps are often called
regular or restricted. We prefer to use the name "classical", in
a sense, that they originate from the continuous deformation of the
finite dimensional irreps of the classical algebra.
In what follows, we shall concentrate on the nilpotent representations.

The method of section 1 for constructing the integrable vertex models
associated to irreps
of quantum groups, can, automatically, be applied to the nilpotent ones.
These representations, that we shall denote by $\pi_{\lambda}$, are given
in a basis $ \{e_r\}^{N'-1}_0$ by:

\begin{eqnarray}
& \pi_{\lambda}(E) e_r = d_{r-1}(\lambda) e_{r-1}  & \nonumber\\
& \pi_{\lambda}(F) e_r = d_{r-1}(\lambda) e_{r+1} & \\
\label{3.2}
& \pi_{\lambda}(K) e_r =  \lambda q^{-2r}  e_{r} & \nonumber
\end{eqnarray}

\noindent
where

\begin{eqnarray}
& d^2_r( \lambda) = [ r+1]  \frac{ \lambda q^{-r} -
\lambda^{-1} q^r}{ q- q^{-1}}& \\
\label{3.3}
& [ x ] = \frac{ q^x - q^{-x}}{q- q^{-1}}& \nonumber
\end{eqnarray}

The eigenvalue of the central element $K^{N'}$ is $\lambda^{N'}$.
The representation (\ref{3.2}) has been chosen because the raising
operator is the transpose of the lowering one, i.e.
$ (\pi_{\lambda}(E))^t = \pi_{\lambda}(F)$. The complex parameter
$\lambda \in C$ labels the nilpotent irreps. It is convenient
to introduce a "generalized" spin $s$ by the relation
\footnote{ Throughout this paper we use $j$ and $s$ to designate "regular"
and nilpotent spins, respectively.}:

\begin{equation}
\lambda = q^{2s} , \; \; 2s = 2s \;\; mod N
\label{3.4}
\end{equation}

For generic values of $\lambda$, the representation (\ref{3.2})
is irreducible. There are, however, a finite collection of $\lambda's$
for which the representation is reducible and fully reducible into
the direct sum of two classical irreps. This happens for the
following values of $\lambda$:

\begin{equation}
\lambda^{(\pm)}_m = \pm q^m  , \; \, m= 0,1,\cdots, N'-2
\label{3.5}
\end{equation}

\noindent
for which
\begin{equation}
\pi_{\lambda^{(\pm)}_m} = \pi^{\pm}_{m/2} \bigoplus
\pi^{\pm}_{(N'-m-2)/2}
\label{3.6}
\end{equation}

We will call the values (\ref{3.5}) orbifold points. Note that they are
fixed points under the action of the coadjoint group defined
in \cite{18}.

Next we want to study certain automorphisms acting on the space of the
nilpotent irreps. Let us introduce two discrete transformations
called spin-parity (P) and charge conjugation or spin-flip (C)
as follows:

\begin{eqnarray}
\lambda^P = - \lambda &, & \lambda^C = q^{-2} \lambda^{-1}
\label{3.7}
\end{eqnarray}

{}From (\ref{3.2}) one derives the relation between the corresponding
irreps:

\begin{eqnarray}
\pi_{\lambda^P}(E) = i \pi_{\lambda}(E)\;, &
\pi_{\lambda^C}(E) ={ \cal{C}} \pi_{\lambda}(F){ \cal{C}} \nonumber\\
\pi_{\lambda^P}(F) = i \pi_{\lambda}(F)\;, &
\pi_{\lambda^C}(F) ={ \cal{C}}\pi_{\lambda}(E) {\cal{C}}  \\
\label{3.8}
\pi_{\lambda^P}(K) = - \pi_{\lambda}(K)\;, &
\pi_{\lambda^C}(K) ={ \cal{C}} \pi_{\lambda}(K^{-1}){ \cal{C}} \nonumber
\end{eqnarray}

\noindent
The charge conjugation matrix $\cal{C}$ is defined by:

\begin{equation}
{\cal{C}} \;\;(e_r) = e_{N'-r-1}
\label{3.9}
\end{equation}

The proof of (\ref{3.8}) is based on the relation $d_r(\lambda^C)=$
$d_{N'-r-2}(\lambda)$. A representation $\pi_{\lambda}$ is called
hermitian if \cite{24}:

\begin{equation}
\pi_{\lambda}(E)^{\dagger} = {v}_{\lambda} \pi_{\lambda}(F)
\; , \; \pi_{\lambda}(K)^{\dagger} =  \pi_{\lambda}(K^{-1})
\label{3.10}
\end{equation}

\noindent
where ${v}_{\lambda} = \pm1$ is a spin-parity of the irrep.
Equation (\ref{3.10}) implies that $\lambda$ is a phase (i.e. $s$
is real) and that the spin parity depends on the spin $s$ as:

\begin{equation}
{v}_s \equiv {v}_{\lambda} = sign \left( \frac{sin 2 \gamma s}{
sin \gamma} \right)
\label{3.11}
\end{equation}

In general, equation (\ref{3.10}) implies certain restrictions on the
allowed values of $\lambda$ :

\begin{eqnarray}
{v}_{\lambda} \;\;d^2_r( \lambda) > 0& for& r=0,1, \dots , N'-2
\label{3.12}
\end{eqnarray}

\noindent
or equivalently
\begin{eqnarray}
{v}_s\;\; sin \gamma k \; sin \gamma ( 2s+1 -k) > 0& for& k=1, \dots
,N'-1
\label{3.13}
\end{eqnarray}

Hence, for a hermitian irrep the sign of $d^2_r(\lambda)$ is
independent of $r$. Our convention for taking the square root is:
\begin{eqnarray}
d_r(\lambda) & =& \left\{ \begin{array}{ll}
[r+1] \left( \frac{1}{[r+1]} \frac{\lambda q^{- r} - \lambda^{-1}
q^r}{q - q^{-1}} \right)^{1/2} & v_{\lambda} = 1 \\
\label{3.14}
i [r+1] \left(- \frac{1}{[r+1]} \frac{\lambda q^{- r} - \lambda^{-1}
q^r}{q - q^{-1}} \right)^{1/2} & v_{\lambda} = - 1
\end{array}
\right.
\end{eqnarray}

A hermitian representation $\pi_{\lambda}$ of $U_q(Sl(2))$, when lifted
to a representation $\pi_{\lambda,u}$ of $U_q( \widehat{ Sl(2)})$,
satisfies the following relations:

\begin{eqnarray}
\left[ \pi_{\lambda,u}(E_0) \right]^{\dagger} & = &  v_{\lambda}
\;\;  \pi_{\lambda,u}(E_1) \nonumber \\
\left[ \pi_{\lambda,u}(F_0) \right]^{\dagger} & = & v_{\lambda}
\;\;  \pi_{\lambda,u}(F_1)  \\
\label{3.04}
\left[ \pi_{\lambda,u}(K_0) \right]^{\dagger} & = & \pi_{\lambda,u}(K_1)
\nonumber
\end{eqnarray}

\noindent
where we have assumed that $u$ is real. \\
These relations are preserved by the comultiplication. The behaviour
of the spin-parity under the discrete symmetries (\ref{3.7}) is:

\begin{eqnarray}
{v}_{\lambda^P} = - {v}_{\lambda}\;,&&
{v}_{\lambda^C} = {v}_{\lambda}
\label{3.15}
\end{eqnarray}

{}From (\ref{3.13}) we deduce that the allowed values of
the spin $s$ belong to  open intervals whose boundaries are
the orbifold points (\ref{3.6}). From now on, for the sake of simplicity,
we deal only with cases where $q = e^{2 \pi i/N}$. The hermiticity
intervals for the cases of $N$ even and odd are given in figure 1.
We observe that the P-operation interchanges the two spin-parity
intervals (i.e.  $ 2s \leftrightarrow 2s + N/2$), while the
C-operation maps each spin-parity interval onto itself leaving invariant
the middle point (i.e. $2s \rightarrow N-2s-2$). In fact, the
middle points of these intervals correspond to classical higher
spin irreps whose spin $j$ is commensurable with the denominator of
the anisotropy, as in equations (\ref{2.21}).
As can be seen from table 1, these irreps are both classical and
nilpotent. The boundaries of the hermiticity regions for $q = e^{2
\pi i/N}$ are the following orbifold points:

\begin{eqnarray}
N\;even :& \lambda = & \pm1 , \pm q^{\frac{ N }{2}-2 } \nonumber \\
N\;odd :& \lambda = & \pm q^{\frac{ N -3}{2}}, \pm q^{\frac{ N -1}{2}}
\label{3.17}
\end{eqnarray}

\noindent
which together with (\ref{3.6}) impliy that at the orbifold points
the hermitian nilpotent irreps break according to the pattern
$ 1 \oplus \frac{N}{2}$ for $N$ even and $\frac{N-1}{2} \oplus
\frac{N+1}{2}$ for $N$ odd. In appendix A we study, in full generality,
the hermiticity regions associated with generic $q$-primitive root
of unity.

\section{q-Chains Based on Nilpotent Irreps}

In this section, we want to find the intertwiner $R$-matrices
associated to the tensor product $ \pi_{\lambda_1,u_1} \otimes
\pi_{\lambda_2,u_2}$. These $R$-matrices can be found, using a recursive
method \cite{25} for solving the intertwiner equations (\ref{2.4}) in the
case of nilpotent irreps. Taking into account the normalization condition

\begin{equation}
R^{0 0}_{0 0} = 1
\label{3.18}
\end{equation}

we have \cite{21}:

\[
R^{\lambda_1 \lambda_2} (u)^{l, r_1 + r_2 - l}_{r_1,r_2} =
\frac{1}{\prod^{r_1+r_2-1}_{j=0} (e^u \lambda_1 \lambda_2
q^{-j} - e^{-u} q^j)} \times
\]

\[
\times {\sum^{r_1}_{l_1 = 0}
\sum^{r_2}_{l_2 = 0}} \left[ \begin{array}{c} r_1
\\ l_1 \end{array} \right] \left[ \begin{array}{c} r_2 \\ l_2
\end{array} \right] \frac{[l] ! [r_2 - l_2] !}{[r_1 + l_2] !
[r_2]!} (q - q^{-1})^{r_1 - l_1 +l_2}
\]

\[
\times \prod^{r_1 + l_2 -1}_{j= r_1} d_j (\lambda_1) \prod^{r_1
+l_2-1}_{j=l_1 +l_2} d_j (\lambda_1) \prod^{r_2
- 1}_{j=r_2 - l_2} d_j (\lambda_2) \prod^{r_1+r_2
-l-1}_{j=r_2 -l_2}d_j(\lambda_2)
\]

\begin{equation}\times \lambda^{l_2}_1 \lambda^{r_1-l_1}_2
\prod^{r_2-l_2-1}_{j=0}
(e^u \lambda_2 q^{-j} - e^{-u} \lambda_1 q^j)
\prod^{l_1 -1}_{j=0} (e^u \lambda_1 q^{-j+r_2-l_2}
-e^{-u} \lambda_2 q^{j+l_2 -r_2})
\label{3.19}
\end{equation}

This is another trigonometric solution to the Yang-Baxter equation.
Strictly speaking, solution (\ref{3.19}) comes with a factor
$q^{l(r_1+ r_2-l) -r_1 r_2}$ which can be eliminated under
an appropriated rescaling of the basis. Next, we list some useful properties
of the $R$-matrix (\ref{3.19}):

i) Normalization

\begin{equation}
R_{\lambda,\lambda}(u=0) = P
\label{3.22}
\end{equation}

where $P$ is the permutation operator $P(v\otimes w)=w\otimes v$,
$v,w\in C^{N'}$.

ii) Unitarity

\begin{equation}
R_{\lambda_1,\lambda_2}(u) R_{\lambda_1,\lambda_2}(-u) = 1
\label{3.23}
\end{equation}

iii) Parity

\begin{equation}
P \; R_{\lambda_1,\lambda_2}(u) P = R_{\lambda_2,\lambda_1}(u)
\label{3.24}
\end{equation}

iv) Spin-Parity

\begin{equation}
R_{\lambda^P_1,\lambda^P_2}(u)
= (1 \otimes e^{i \pi {\cal N}}) R_{\lambda_1,\lambda_2}(u)
( e^{i \pi {\cal N}} \oplus 1 )
\label{3.25}
\end{equation}

where ${\cal N}$ is the number operator:

\begin{equation}
{\cal N} e_r = r e_r
\label{3.26}
\end{equation}

v) Charge-Conjugation

\begin{equation}
R_{\lambda^C_1,\lambda^C_2}(u) =
\frac{1}{R^{N'-1, N'-1}_{N'-1,N'-1}(\lambda_1,\lambda_2;u)}
( {\cal C} \otimes {\cal C} ) R_{\lambda_1,\lambda_2}(u)
( {\cal C} \otimes {\cal C} )
\label{3.27}
\end{equation}

vi) Hermiticity

\begin{equation}
R_{\lambda_1,\lambda_2}^{\dagger}(u) = R_{\lambda_1,\lambda_2}(-u)
\label{3.28}
\end{equation}

This last equation holds true when both irreps $\lambda_1$ and $\lambda_2$
are hermitian and have the same spin parity. It reflects the fact
that the dagger operation is compatible with the comultiplication
of $U_q({\widehat Sl(2)})$, namely:

\begin{eqnarray}
\left[ \pi_{\lambda_1,u_1} \otimes  \pi_{\lambda_2,u_2}
(\Delta (E_0)) \right]^{\dagger} & =&
{v}_{s_1}  \;\;\; \pi_{\lambda_1,u_1} \otimes  \pi_{\lambda_2,u_2}
( \Delta (E_1))\nonumber \\
\left[ \pi_{\lambda_1,u_1} \otimes  \pi_{\lambda_2,u_2}
(\Delta (F_0)) \right]^{\dagger} & = &
{v}_{s_1} \;\;\;  \pi_{\lambda_1,u_1} \otimes  \pi_{\lambda_2,u_2}
( \Delta (F_1))
\label{3.29}
\end{eqnarray}

Using the $R$-matrix (\ref{3.19}) we can define an integrable
vertex model whose integrability condition (\ref{2.8}) reads:

\begin{equation}
[ t^{\lambda_1, \lambda}(u_1),t^{\lambda_2, \lambda}(u_2)]=0
\label{3.30}
\end{equation}

In this case, the Bethe equations are a generalization of (\ref{2.18})
to continuous values of the spin:

\begin{equation}
\left( \frac{sinh \frac{\gamma}{2} (u_j + 2s i)}{
sinh \frac{\gamma}{2} (u_j - 2s i)} \right)^L=
\prod^M_{k=1,k \neq j}
 \frac{sinh \frac{\gamma}{2} (u_j - u_k +2i)}{
sinh \frac{\gamma}{2} (u_j - u_k - 2i)}
\label{3.31}
\end{equation}

The spin-chain hamiltonian (\ref{2.19}) will, of course, depend on the
spin $s$:

\begin{equation}
H_{\lambda} \equiv H_{s}(\gamma) =
i I \frac{\partial}{\partial u} log \;\; t^{\lambda,\lambda}(u)|_{u=0}
\label{3.32}
\end{equation}

\noindent
where $I$ is an overall coupling constant.

Before giving the general form of these hamiltonians, it is worth
to exhibit their properties which follow from those of the $R$-matrix:

i) Hermiticity: If  $\pi_{\lambda} $ is hermitian then

\begin{equation}
H^{\dagger}_{\lambda} = H_{\lambda}
\label{3.33}
\end{equation}

ii) Parity
\begin{equation}
P_L H_{\lambda} P_L = H_{\lambda}
\label{3.34}
\end{equation}

where $P_L$ is the operator that reverses the order of the sites along
the chain.

iii) Spin-Parity
\begin{equation}
H_{\lambda^P} = \Omega H_{\lambda} \Omega^{\dagger}
\label{3.36}
\end{equation}

where $ \Omega = \prod^L_{j=1} e^{i \pi j {\cal N}_j}$.

iv) Charge-Conjugation

\begin{equation}
H_{\lambda^C} = V H_{\lambda} V - i L \frac{\partial}{\partial u}
R^{ N' -1 N' -1}_{N' -1 N' -1}(\lambda,\lambda;u)|_{u=0}
\label{3.37}
\end{equation}

where $V = \prod^L_{j=1} {\cal C}_j$.
\\
Since $\Omega$ and $V$ are unitary operators, we conclude that the
hamiltonians associated to $\lambda, \lambda_P$ and $\lambda_C$
describe the same physics. Hence, we shall restrict ourselves to
a fundamental region  of the P and C operations
in the $\lambda $ space. For any value of $q$ we can choose the
half intervals going from the classical point up to the orbifold point
within a hermiticity region. In the cases where $q= e^{2 \pi i/N}$,
these intervals are given by:

\begin{eqnarray}
N\; even: & & j \leq s < j + 1/2 \nonumber \\
N\; odd: & & j \leq s < j + 1/4
\label{3.38}
\end{eqnarray}

Thus, we have a one parameter family of Hamiltonians $H_s(\gamma$
$= 2 \pi/N)$ on the interval (\ref{3.38}), which we can compare with
the higher spin $XXZ$ hamiltonians $ H_j (\gamma)$ of section 1. The
results of this comparison are graphically presented in figure 2.
Along the line I, one continuously modifies the anisotropy $\gamma$
keeping the spin $j$ fixed, while along the line II, one varies the "spin".
These two lines meet at a single point $A$
which corresponds exactly to the commensurable higher spin cases of
(\ref{2.21}).

A peculiar feature of the nilpotent models (i.e. line II)
is the existence of another local conserved quantity obtained
by taking the derivative of the transfer matrix with respect to
$\lambda_1$ :

\begin{equation}
Q_{\lambda} = 2 \lambda_1 \frac{\partial}{\partial \lambda_1}
log t^{\lambda_1, \lambda}(u=0) |_{\lambda_1 = \lambda}
\label{3.39}
\end{equation}

The counterparts of (\ref{3.23} - \ref{3.28}) for
$Q_{\lambda}$ are:

i) Hermiticity

\begin{equation}
Q^{\dagger}_{\lambda} = Q_{\lambda}
\label{3.41}
\end{equation}

ii) Parity

\begin{equation}
P_L Q_{\lambda} P_L = - Q_{\lambda}
\label{3.42}
\end{equation}

iii) Spin-Parity

\begin{equation}
Q_{\lambda^P} =  - \Omega Q_{\lambda} \Omega^{\dagger}
\label{3.43}
\end{equation}

iv) Charge-Conjugation

\begin{equation}
Q_{\lambda^C} = -V Q_{\lambda} V
\label{3.44}
\end{equation}

Next, we shall consider a few examples in order to get some insight
into the properties of the models we intend to study.

\section*{Case N=4}

The point $A$ of figure 2 has $\gamma = \pi/2$ and $j=1/2$ and
corresponds simply to the $XX$ model:

\begin{equation}
H_{XX} =
\sum^L_{i=1}( \sigma^X_i \sigma^X_{i+1} +
\sigma^Y_i \sigma^Y_{i+1} )
\label{3.45}
\end{equation}

\noindent
which is  equivalent, after a Jordan-Wigner transform,
to a free fermion model. Line I of figure 2 corresponds to the $XXZ$ models
for different values of the anisotropy $\gamma$ . The hamiltonians on line II
are:

\begin{equation}
H_{\lambda} = \frac{i}{\lambda - {\lambda}^{-1}}
\sum^L_{i=1}( \sigma^X_i \sigma^X_{i+1} +
\sigma^Y_i \sigma^Y_{i+1} + \frac{( \lambda + \lambda^{-1})}{2}
(\sigma^Z_i+\sigma^Z_{i+1}))
\label{3.46}
\end{equation}

\noindent
and can be recognized as those of the $XX$ models with an external magnetic
field given by $ \frac{\lambda + \lambda^{-1}}{2} = cos \pi s$.
The hermiticity region of (\ref{3.46}) is $2s \in (0,2)$ whose
boundaries correspond to the critical values of the magnetic field.
The operator $Q$ is given by:

\begin{equation}
Q_{\lambda} = \frac{-i}{\lambda - {\lambda}^{-1}}
\sum^L_{i=1}( \sigma^X_i \sigma^Y_{i+1} -
\sigma^Y_i \sigma^X_{i+1})
\label{3.47}
\end{equation}

This is a hopping hamiltonian which is parity odd in contrast with
(\ref{3.46}) which is parity even. Combining these two commuting
operators we can construct a hopping hamiltonian with a complex hopping
parameter. The hamiltonian (\ref{3.46}) generalizes the free
fermion model in a quantum group symmetric way (the corresponding $R$-matrix
was first introduced \cite{26} from somewhat different point of view).
In fact, the nilpotent $R$-matrix coincides, in this case, with the $R$-matrix
of $Gl(1,1)_q$ ($q = \lambda)$ in the fundamental representation
\cite{27}.

\section*{Case N=3}

The hamiltonians on line I were constructed by
Fateev and Zamolodchikov and are associated with the spin 1
$XXZ$ models \cite{3}:

\[
H_{FZ}(q) = \sum^{L}_{k = 1}
S^X_k S^X_{k+1} + S^Y_k S^Y_{k + 1}
+ \frac{ q^2 + {q}^{-2}}{2}
S^Z_k  S^Z_{k + 1}
\]

\begin{equation}
-( S^X_k S^X_{k+1} + S^Y_k S^Y_{k + 1} )^2
- \frac{ q^2 + {q}^{-2}}{2}
(S^Z_k  S^Z_{k + 1})^2 \\
\label{3.48}
\end{equation}

\[
+ (1 - q - q^{-1})
\left[  ( S^X_k S^X_{k+1} + S^Y_k S^Y_{k + 1} )S^Z_k S^Z_{k + 1}
+ \leftrightarrow \right]
\]

\[
+ \frac{ q^2 + {q}^{-2}- 2}{2}
[(S^Z_k)^2 +(  S^Z_{k + 1})^2 ]
\]

\noindent
where $S^X, S^Y $ and $S^Z$ are the classical spin 1 matrices.
At the isotropic point $\gamma =0$, (\ref{3.48}) becomes the
hamiltonian (\ref{2.20}).

The later model is gapless and corresponds to a
$WZW$ model $\widehat{ SU(2)} $ with level $k = 2j = 2$
which has central extension $c=3/2$ \cite{8}.
The hamiltonians on line II can be written as:

\[
H({\lambda}) = - \frac{1}{ sin \gamma \;\;d_0^2 d_1^2}
\sum^{L}_{k=1}
\frac{{\lambda} q+ {\lambda}^{-1} q^{-1}}{2}
( S^X_k S^X_{k+1} + S^Y_k S^Y_{k + 1} ) - \frac{1}{2}
S^Z_k S^Z_{k + 1}
\]

\[
- ( S^X_k S^X_{k+1} + S^Y_k S^Y_{k + 1} )^2 + \frac{1}{2}
(S^Z_k S^Z_{k + 1})^2
\]

\begin{equation}
+ \left( \frac {{\lambda} q + {\lambda}^{-1} q^{-1}}{2}
+d_0d_1 \right) \;
\left[  ( S^X_k S^X_{k+1} + S^Y_k S^Y_{k + 1} )S^Z_k S^Z_{k + 1}
+ \leftrightarrow \right]\\
\label{3.50}
\end{equation}

\[
- \frac{3}{2}
\left( (S^Z_k)^2 + ( S^Z_{k + 1})^2 \right)
- \frac {{\lambda} q - {\lambda}^{-1} q^{-1}}
{2 (q - q^{-1}) }
( S^X_k S^X_{k+1} + S^Y_k S^Y_{k + 1} )(S^Z_k +S^Z_{k + 1})
\]

\[
+ \frac {{\lambda}^2 q^{-1} - {\lambda}^{-2}
q}
{2 (q - q^{-1}) }
\left( S^Z_k +S^Z_{k + 1} \right)\\
\]

\noindent
which coincides with (\ref{3.48}) at point $A$ , i.e. $\lambda= q^2$
up to an overall factor. From this result we see that the hamiltonian
$H_{FZ}(q)$ at $q^3 = 1$ possesses an additional symmetry given by the
operator $Q_{\lambda}$ with $\lambda= q^2$.

In the $N=4$ case both hamiltonians  $H_{\lambda} $ and
$ Q_{\lambda}$ were seen to describe the hopping of a free fermion
in the presence of a magnetic field. In the $N=3$ case one has
three degrees of freedom per site. This fact can be accommodated
in a model of two fermions $A$ and $B$ subject to the constraint of no
double occupancy. Upon the identification $e_0 \rightarrow A,
e_1 \rightarrow \emptyset, e_2 \rightarrow B$, where $\emptyset$
denotes an empty site, we deduce that the hamiltonian (\ref{3.50}) at
$\lambda = q^2$ describes the following processes in the terminology
of the ref. \cite{28}:

1) Diffusion

\begin{eqnarray}
A+\emptyset \rightarrow \emptyset + A, &
B+\emptyset \rightarrow \emptyset + B, &
A+B \rightarrow B+ A \nonumber \\
\emptyset + A \rightarrow  A +\emptyset,  &
\emptyset + B \rightarrow  B +\emptyset, &
B + A \rightarrow  A + B \nonumber \\
H^{\emptyset A}_{A \emptyset } = H^{A \emptyset }_{\emptyset A}= &
H^{\emptyset B}_{B \emptyset } = H^{B \emptyset }_{\emptyset B}= &
-H^{B A}_{A B } = -H^{A B }_{B A} =-\frac{1}{sin \gamma}
\label{3.52}
\end{eqnarray}

where $\gamma =2 \pi/3$.
\\
2) Annihilation
\begin{eqnarray}
A + B &  \rightarrow  & \emptyset + \emptyset \nonumber \\
B + A &  \rightarrow  & \emptyset + \emptyset \nonumber \\
H^{\emptyset \emptyset}_{A B} =& H_{B A}^{\emptyset \emptyset}
& = -\frac{1}{sin \gamma}
\label{3.53}
\end{eqnarray}

3) Creation
\begin{eqnarray}
\emptyset + \emptyset & \rightarrow & A + B \nonumber \\
\emptyset + \emptyset & \rightarrow & B + A \nonumber \\
H^{A B}_{\emptyset \emptyset} = & H^{B A}_{\emptyset \emptyset}=&
- \frac{1}{ sin \gamma}
\label{3.54}
\end{eqnarray}

The hamiltonian has in addition a chemical potential term
$ \mu \otimes 1 + 1 \otimes \mu$ with :

\begin{eqnarray}
\mu_A = \mu_B =  \frac{1}{2 sin \gamma}& & \mu_{\emptyset}=0
\label{3.55}
\end{eqnarray}

We obtained a "chemical interpretation" along the lines
of ref. \cite{28} of the hamiltonian $H_{\lambda= q^2} = H_{FZ}(q^3=1)$.

Another interesting hamiltonian emerges when one reaches an
orbifold point. As can be seen from (\ref{3.50}), $H_{\lambda}$ has
a simple pole at an orbifold point $\lambda_0$. Then, it makes sense to define
an orbifold hamiltonian $H^{(orb)}_{\lambda_0}$ as a residue
of $H_{\lambda}$ at this point, namely:

\begin{equation}
H^{(orb)}_{\lambda_0}=
\lim_{\lambda \rightarrow \lambda_0} \prod^{N' -2}_{r=0}
d^2_r ( \lambda) H_{\lambda}
\label{3.56}
\end{equation}

Choosing $\lambda_0 = 1$ we get the following "orbifold" hamiltonian:

\begin{equation}
H^{(orb)}_{\lambda=1} = \frac{1}{sin \gamma}
\sum^L_{j=1} \left( \tau^+_j \tau^-_{j+1} + \tau^-_j \tau^+_{j+1}
+\rho^+_j \rho^-_{j+1} + \rho^-_j \rho^+_ {j+1} - \sigma^0_j -
\sigma^0_{j+1} \right)
\label{3.57}
\end{equation}

\noindent
where $\tau,\rho,\sigma$ are different embeddings of Pauli matrices acting
in $C^3$:
\begin{eqnarray}
\tau^+ = E^{12}, & \rho^+ = E^{13},& \sigma^0 = E^{22}+
E^{33}\nonumber \\
\tau^- =E^{21}, & \rho^- = E^{31}&
\label{3.58}
\end{eqnarray}

\noindent
and $E^{i j}$ is the $3 \times 3$ matrix with 1 in the $i-j $ position
and zeros elsewhere.
Eqn. (\ref{3.57}) has a classical $SU(2) \times U(1)$ invariance generated
by $ \sum_j \sigma^{\pm}_j, \sum_j \sigma^Z , \sum_j \sigma^0$.
In ref. \cite{28} this hamiltonian was given a chemical interpretation in
terms of diffusion processes only.
Another possible interpretation of (\ref{3.57}) is a Hubbard model
with infinite Coulomb repulsion or, equivalently, a $t-J$ model with
zero spin-spin coupling. It is known that these models have an infinitely
degenerated ground state when $L\rightarrow\infty$. This degeneracy is broken
in the $t-J$ model
by the spin-spin coupling while, in our case, it is broken by the
values of $\lambda$ different from 1. We will further discuss $N=3$ case in
section 5.

\section*{Generic N Case}

Taking into account the complicated structure of the hamiltonian
(\ref{3.50}) for the $N=3$ case, it would seem hopeless to find an explicit
expression of the hamiltonian for the generic values of $N$. Fortunately,
this is not so and, in fact, we get the following expression:

\begin{equation}
H_{\lambda}= -\sum^L_{j=1} \sum^{N'-1}_{n=1}
\left( \frac{1}{sin \gamma n} \left[ \left( \Sigma^+_j(\lambda)
\Sigma^-_{j+1}(\lambda) \right)^n + \left( \Sigma^-_j (\lambda)
\Sigma^+_{j+1} (\lambda) \right)^n \right] + \mu_j(\lambda)
+ \mu_{j+1}(\lambda) \right)
\label{3.60}
\end{equation}

\noindent
where the matrices $\Sigma^{\pm}(\lambda)$ are given by:

\begin{eqnarray}
\Sigma^+(\lambda) e_r & =& \left( \frac{q^r - q^{-r}}{q^{r-1} \lambda^{-1}
- q^{-r +1} \lambda} \right)^{1/2} \; e_{r-1} \nonumber \\
\Sigma^- (\lambda) & = & \left( \Sigma^+ (\lambda) \right)^{\dag}
\label{3.61}
\end{eqnarray}

\noindent
and
\begin{eqnarray}
\mu(\lambda) \; e_0 & =& 0 \nonumber \\
\mu(\lambda) \; e_r &=& - i \sum^{r-1}_{j=0}
\frac{ \lambda^{-1} q^j + \lambda q^{-j}}{\lambda^{-1} q^j-  \lambda q^{-j}}
\;\; e_r
\label{3.62}
\end{eqnarray}

\noindent
In deriving (\ref{3.60}), we have assumed that $\lambda$ has a negative
parity. \\
The hamiltonian (\ref{3.60}) exhibits strong formal similarities
with the one of the Chiral Potts models \cite{19}:

\begin{equation}
H_{CP} = -\sum^L_{j=1} \sum^{N-1}_{n=1}
\left[ \alpha_n \left( X_j X^{\dag}_{j+1} \right)^n
+ \bar{\alpha}_n Z^n_j \right]
\label{3.63}
\end{equation}

\noindent
where the coupling constants $\alpha_n$ and $\bar{\alpha}_n$ are
given by:

\begin{eqnarray}
\alpha_k = \frac{e^{i \frac{2k-N}{N} \phi} }{sin( \frac{\pi k}{N})},&
\bar{\alpha}_k =  \frac{ k' e^{i \frac{2k-N}{N} \bar{\phi}} }
{sin( \frac{\pi k}{N})},&
cos \phi = k'cos \bar{\phi}
\label{3.64}
\end{eqnarray}

\noindent
and $X$ and $Z$ satisfy $X^N = Z^N =1$, $X Z = e^{2 \pi i /N} Z X$.
A possible basis for $X$ and $Z$ is:
\begin{eqnarray}
X e_r &=&  e_{r-1} \;\;modN \nonumber \\
Z e_r &=& e^{2 \pi i r/N} e_r
\label{3.65}
\end{eqnarray}

To bring closer the relationship between (\ref{3.60}) and (\ref{3.63})
we will consider the case $\lambda= q^{-1}$ and $\gamma = \pi/N$, namely:

\begin{equation}
H_{\lambda =q^{-1}}(q = e^{i \pi /N})
= -\sum^L_{j=1} \sum^{N-1}_{n=1}
\left( \frac{1}{sin ( \frac{\pi n}{N})} \left[ \left( \Sigma^+_j
\Sigma^-_{j+1} \right)^n + \left( \Sigma^-_j
\Sigma^+_{j+1} \right)^n \right] +
( 1- \frac{2n}{N}) e^{i \pi n/N} Z^n_j \right)
\label{3.66}
\end{equation}

\noindent
where $\Sigma^{\pm}\equiv \Sigma^{\pm}(\lambda = q^{-1})$ are an
$N$-dimensional version of the Pauli matrices $\sigma^{\pm}$ (indeed,
$\Sigma^{\pm}= \sigma^{\pm}$ for $N= 2$).
The origin of the formal similarity between (\ref{3.66})
and the  Chiral Potts hamiltonians is due to the chain of reductions
cyclic $\rightarrow$ semicyclic $\rightarrow$ nilpotent  irreps
and the fact that the chiral Potts model is built up from the cyclic irreps.
Indeed, if we choose $\phi=0$ and $\bar{\phi} \rightarrow
\pi/2$ in (\ref{3.63}) we get the same $Z$ terms up to a divergent piece
proportional to the number operator. Still, one has to explain why
the cyclic operator $X$ reduces to the nilpotent operator
$\Sigma^+$. Although there are formal similarities between the Chiral Potts
model and the one discussed here, we believe that these two models are
not equivalent. The main reason is that Chiral Potts preserves a $Z_N$
symmetry which is broken in our case.

The discussion above suggests that the models we constructed
belong to a more general class of integrable theories, whose
hamiltonians have the generic structure:

\begin{equation}
H_{CN}
= -\sum^L_{j=1} \sum^{N'-1}_{n=1}
\left[ t_n \left( \Sigma^+_j
\Sigma^-_{j+1} \right)^n + \bar{t}_n \left( \Sigma^-_j
\Sigma^+_{j+1} \right)^n +
h_n Z^n_j \right]
\label{3.68}
\end{equation}

This kind of hamiltonians can be obtained through the extension of the
Hopf algebra $U_q(\widehat{Sl(2)})$ by
a central element $z$ with non trivial comultiplications.
In this way, the parity is broken and one gets chiral hamiltonians.

To end this section, we want to analyze what happens when
we reach an orbifold point of the hamiltonian (\ref{3.60}).
If $\lambda$ approaches $\pm q^m$, we see that $H^{(orb)}$
has the block diagonal form $H^{I, II}_{II, I} $ or $H^{II, I }_{I, II},$
where I and II denotes the two sectors in which the nilpotent
irrep breaks at the orbifold point. This means that the only
processes taking place are diffusion ones between states which belong to
different sectors $ I + II \leftrightarrow II + I$.

\chapter*{3. Thermodynamics}

\setcounter{chapter}{3}
\setcounter{section}{0}
\setcounter{equation}{0}

\section{Bethe Equations, The String Hypothesis}

In this section we will consider in some details the Bethe equations
(\ref{3.31}) with generic spin $s$ in the hermiticity region (\ref{3.38}).
In the thermodynamic limit $L\rightarrow\infty$, we will assume the string
hypothesis \cite{17}. This means that the solutions to (\ref{3.31}) are
organized into the strings of roots sharing a common real value.
These strings are characterized by length $n$ and parity $v_n$ as
follows:
\begin{equation}
u_l^{(n)}=u+i[n+1-2l+\frac{\pi}{2\gamma}(1-v_sv_n)]\;;\;\;\;\;l=1,...,n
\label{4.1}
\end{equation}

\noindent
where $u$ represents common the real part and $v_s$ is the spin parity
introduced in (\ref{3.11}). Strings of roots can be intuitively interpreted as
bound states of the spin waves. From the Bethe equations (\ref{3.31}) we
derive the consistency conditions for the allowed strings (\ref{4.1}):

\begin{equation}
\left|\prod_{j=1}^k \frac{sh\frac{\gamma}{2}(u_j^{(n)}+2is)}{sh\frac{\gamma}{2}
(u_j^{(n)}-2is)}\right|>1\;;\;\;\;\;k=1,\dots,n-1
\label{4.2}
\end{equation}
\\
\noindent
It is not hard to prove that the consistency condition (\ref{4.2}) for the
Takahashi strings (\ref{4.1}) is equivalent, for asymptotically large
rapidities, to the hermiticity condition (\ref{3.13}) for irreps of dimension
$n$ and spin parity $v_n$. In fact, for $u\rightarrow\infty$ we get from
(\ref{4.2}):

\begin{equation}
v_s \frac{\sin(2\gamma s)}{\sin\gamma}v_n\sin(\gamma n)\sin\gamma(n-k)>0\;;
\;\;\;\;k=1,\dots,n-1 \\
\label{4.3}
\end{equation}

\noindent
Using (\ref{3.11}) we obtain the hermiticity condition (\ref{3.13}) for irreps
of dimension $n$. Notice that the condition (\ref{4.2}), for
$u\rightarrow\infty$, fixes the allowed strings in a form completely
independent of the value of the generalized spin $s$. To feel $s$ we need to
consider (\ref{4.2}) for generic values of the rapidities. We now show that
if $s$ is in the hermiticity region (\ref{3.38}) then the asymptotically
allowed strings (\ref{4.3}) will satisfy (\ref{4.2}) for arbitrary values of
rapidities. To this end we rewrite (\ref{4.2}) as:

\begin{equation}
\prod_{j=1}^k\frac{z-\cos(y+2\gamma(s-j))}{z-\cos(y-2\gamma(s+j))}>1
\label{4.4}
\end{equation}

\noindent
where $z=ch\gamma u$; $y=\gamma(n+1+p_0\frac{1-v_nv_s}{2})$
and $1\leq k \leq (n-1)$. It is straightforward to prove the following
identity:

\begin{equation}
\prod_{j=1}^k\frac{z-\cos(y+2\gamma(s-j))}{z-\cos(y-2\gamma(s+j))}=
1+\sum_{m=0}^{k-1}\prod_{l=0}^m\frac{2v_nv_s\sin\gamma(2s-l)}{\sin\gamma(l+1)}
\frac{\sin\gamma(n-k+l)\sin\gamma(k-l)}{z-v_nv_s\cos\gamma(n+1-2k+2l-2s)}
\label{4.5}
\end{equation}

\noindent
Taking into account the hermiticity condition (\ref{3.13}) and the Takahashi
condition (\ref{4.3}) one immediately concludes that l.h.s. of eqn.
(\ref{4.4}) is, indeed, greater than one.

Making use of "Takahashi zone" terminology \cite{17} (also see appendix A)
we have, for allowed strings ($n_j,v_j$), the following:

\begin{eqnarray}
N & {\rm even}: & \left\{ \begin{array}{lll}
0-{\rm zone} & n_j=j, v_{n_j}=+1 & 1\leq j \leq \nu-1 \\
1-{\rm zone} & n_{\nu}=1, v_{\nu}=-1 & j=\nu \\
\end{array}
\right. \nonumber \\
& & \label{4.6} \\
N & {\rm odd}: & \left\{ \begin{array}{ll}
0-{\rm zone} & \;\;\;\; n_j=j, v_j=+1
\;\;\;\;\;\;\;\;\;\;\;\;\;\;\;\;\;\;\;\;\;\; 1\leq j \leq \nu-1 \\
1-{\rm zone} & \left\{ \begin{array}{ll}
n_{\nu}=1, v_{\nu}=-1 & \;\;\; j=\nu \\
n_{\nu+1}=\nu+1, \; v_{\nu+1}=+1 & \;\;\; j=\nu+1 \\
\end{array}
\right. \nonumber \\
2-{\rm zone} & \;\;\;\; n_{\nu+2}=\nu, \;\;\; v_{\nu+2}=+1
\;\;\;\;\;\;\;\;\;\;\; j=\nu+2 \\
\end{array}
\right. \nonumber
\end{eqnarray}

\noindent
where $\nu=\frac{N}{2}(\frac{N-1}{2})$ for $N$-even (odd).

In the thermodynamic limit, equations (\ref{3.31}) become:

\begin{equation}
a_j(u)=(-1)^{r_j}(\rho_j+\rho_j^{(h)})+\sum_kT_{jk}\ast\rho_k
\label{4.7}
\end{equation}

\noindent
where $\rho_j(\rho_j^{(h)})$ is density of $j$-strings ($j$-holes) and
$(-1)^{r_j}=sign(a_j)$. The Fourier images of the functions which appear in
the equation above are given by:

\begin{eqnarray}
\hat{T}_{jk} & = & g(w;|n_j-n_k|;v_{n_j}v_{n_k}) +
g(w;|n_j+n_k|,v_{n_j}v_{n_k})+  \nonumber \\
&  & + 2 (1-\delta_{1,min(n_j,n_k)}) \sum_{l=1}^{min(n_j,n_k)-1}
g(w;|n_j-n_k|+2l;v_{n_j}v_{n_k})
\label{4.8} \\
\hat{a_j} & = & \sum_{l=0}^{n_j-1}g(w;2s+1-n_j+2l,v_sv_{n_j}) \nonumber \\
g(w;n;v) & = & - \frac{sh2p_0 w ((\frac{n}{2p_0}+\frac{1-v}{4}))}
{shp_0 w};
\;\;\;\;\;\;((x)) \;\;\; {\rm is\;\;Dedekind\;\;function} \nonumber
\end{eqnarray}
\\
\noindent
The "bare" energy of strings of the type $j$ is given by:

\begin{equation}
E_{n_j}(u)=-\frac{4\pi}{\gamma}Ia_j(u)
\label{4.9}
\end{equation}

\noindent
where $I$ is an overall coupling constant which is assumed to be positive
throughout this paper.

Clearly, the equation (\ref{4.7}) makes sense only if the sign of
$a_j(u)$ (or $E_{n_j}(u)$) does not change when $u$ varies from zero to
infinity. It is a bit surprising that no proof of this fact has been given in
the literature. The following is the proof that, indeed, there is no sign
crossing for $a_j(u)$ as long as $n_j$ is an allowed Takahashi string length
and $s$ belongs to the hermiticity region (\ref{3.38}).
We represent energy $E_n$ in the following form:

\begin{eqnarray}
E_n(z)=-\frac{2Iv_nv_s\sin(2\gamma s)\sin(\gamma n)}
{\sin\gamma[z-v_nv_s\cos\gamma(2s+n-1)]}\times \nonumber \\
\times\{1+\sum_{m=1}^{n-1}\prod_{l=1}^m \frac{2v_nv_s\sin\gamma(2s-l)
\sin\gamma(n-l)\sin\gamma l}{\sin\gamma(l+1)[z-v_nv_s\cos\gamma(2s+n-2l-1)]}\}
\label{4.10}
\end{eqnarray}
\\
\noindent
Once again, making use of formulas (\ref{3.13}) and (\ref{4.3}), one can
infer that the expression inside of the figure brackets in the equation above
is always positive and, therefore, the $sign\;of\;E_n$ is:

\begin{equation}
sign(E_n)=-sign[v_nv_s\sin(2\gamma s)\sin(\gamma n)\sin\gamma]=
-(-1)^{r_j}=-(-1)^i
\label{4.11}
\end{equation}

\noindent
with $i$ being a label of the "Takahashi zone" where $n_j$ string lives.

Before we move on, let us make the following comment. Irregular spin $s$, in
some sense, plays the role of the "internal" magnetic field, producing, for
instance, the nonvanishing ground state magnetization \cite{21}. However,
this interpretation cannot be taken literary because other attributes of
magnetic field are absent. For example, none of the string energies change
their signs as opposite to the case where system is exposed to the external
magnetic field.

Renormalized or "dressed" energy of the string is defined as:

\begin{equation}
\exp \left[ \beta \epsilon_j(u)\right] = \frac{\rho_j^{(h)}(u)}{\rho_j(u)}
\equiv \eta_j \;;\;\;\;\; \beta=\frac{1}{T}
\label{4.12}
\end{equation}

\noindent
with $T$ being temperature. \\
Equations for the "dressed" energies can be obtained by minimizing the free
energy $F$

\begin{equation}
\frac{F}{L}=-T\sum_j \int_{-\infty}^{+\infty}
du|a_j(u)|
ln(1+\eta_j^{-1})
\label{4.13}
\end{equation}

\noindent
in the presence of the Bethe ansatz constraints (\ref{4.5}) as was first
proposed in ref. \cite{29}. The resulting Dyson like equation is:

\begin{equation}
\ln\eta_j=-4p_0I\beta a_j+\sum_k (-1)^{r_k} T_{jk}\ast\ln (1+\eta_k^{-1})
\label{4.14}
\end{equation}

Before giving the explicit solution to (\ref{4.14}) let us briefly comment on
its physical meaning. The positive energy strings (and the holes in
distribution of negative energy strings) can be interpreted as particle like
excitations.
The negative energy strings will determine the ground state defined as a state
where all negative energy modes are filled. Finally, the zero energy strings
reflect the symmetries of the model and will be used to define the internal
quantum numbers of all elementary excitations. \\
To solve equation (\ref{4.14}), it is convenient first to invert the matrix
$T$ (see appendix B). Then, this equation can be converted to the following
form:

\begin{eqnarray}
{\rm Case\;\;N\;\;even}  & & \nonumber \\
\ln{\eta}_j & = & s_1\ast [\ln(1+\eta_{j+1})(1+\eta_{j-1})+\delta_{j,\nu-2}
\ln(1+\eta_{\nu}^{-1})]\;;\;\;\;1\leq j\leq\nu-2 \nonumber \\
\ln\eta_{\nu-1+\tilde{r}} & = & (-1)^{\tilde{r}}[s_1\ast \ln(1+\eta_{\nu-2})+
\frac{Ip_0}{T}\frac{\sin\pi x}{ch\frac{\pi u}{2}+
(-1)^{\tilde{r}}\cos \pi x}]\;;\;\;\;\tilde{r}\equiv 0,1 \nonumber \\
\nonumber \\
{\rm Case\;\; N\;\; odd} & & \label{4.15} \\
\ln{\eta}_j & = & s_1\ast [\ln(1+\eta_{j+1})(1+\eta_{j-1})]+\delta_{j,\nu-1}
[s_1\ast\ln(1+\eta_j)+ \nonumber \\
& + & (s_2-s_1)\ast\ln(1+\eta_{\nu})-\frac{4p_0I}{T}f(u)] \;;\;\;\;\;
1\leq j\leq\nu-1 \nonumber \\
\ln\eta_{\nu} & = & s_2\ast\ln(1+\eta_{\nu+1})(1+\eta_{\nu+2}^{-1})
(1+\eta_{\nu-1})^{-1} \nonumber \\
\ln\eta_{\nu+1+\tilde{r}} & = & (-1)^{\tilde{r}}[s_2\ast
\ln(1+\eta_{\nu})-\frac{2Ip_0}{T}
\frac{\cos 2\pi x}{ch\pi u+(-1)^{\tilde{r}}\sin 2\pi x}]\;;\;\;\;
\tilde{r}\equiv 0,1 \nonumber
\end{eqnarray}

\noindent
where $s_l(u)=\frac{l}{4}sech\frac{\pi ul}{2}$, $l\equiv 1,2$; $\eta_0=0$;
$x=s-p_0\frac{3-v_s}{4}$; $\hat f(w)=2sh(\frac{w}{2})\frac{chw(2x+1)}{sh2w}$.\\
The equations above can be easily solved in two cases: $u\rightarrow\infty$
(or $\beta\rightarrow 0$) and $u\approx 0$ \cite{17}.
These cases will be important later for the central charge calculations.
Results are summarized in table 2 below:

\begin{center}
\begin{tabular}{|c|c|}
\hline
$N$ even & $N$ odd \\
\hline
$1+\overline{\eta}_j=(j+1)^2\;;\;\;\;\;  1\leq j \leq \nu-2$
& $1+\overline{\eta}_j=(j+1)^2\;;\;\;\;\;1\leq j \leq \nu-1$ \\
$\overline{\eta}_{\nu-1}=\overline{\eta}_{\nu}^{-1}=\nu-1$
& $1+\overline{\eta}_{\nu}=\overline{\eta}_{\nu+1}^2=(1+\nu^{-1})^2$\\
& $\overline{\eta}_{\nu+2}=1-(1+\nu)^{-1}$ \\
\hline
$1+\overline{\overline{\eta}}_j=\frac{\sin^2\frac{\pi}{\nu+1}(j+1)}
{\sin^2\frac{\pi}{\nu+1}}\;;\;\;\;1\leq j\leq \nu-1$
& $1+\overline{\overline{\eta}}_j=\frac{\sin^2\frac{\pi}{\nu+2}(j+1)}
{\sin^2\frac{\pi}{\nu+2}}\;;\;\;\;1\leq j\leq \nu-1$ \\
$\overline{\overline{\eta}}_\nu=\infty$
& $\overline{\overline{\eta}}_\nu=\overline{\overline{\eta}}_{\nu+1}=
\overline{\overline{\eta}}_{\nu+2}^{-1}=\infty$ \\
\hline
\end{tabular}
\end{center}
\begin{center}
Table 2.
\end{center}

\noindent
In table 2 above, $\overline{\eta}=\eta(\infty)$ and
$\overline{\overline{\eta}}=\eta(0)$.

It is relatively simple to convince oneself, using equations (\ref{4.15}),
that all strings are non-negative except $\nu-1$ string for $N$ even and
$\nu+2$ for $N$ odd. Making use of this fact, it is now easy to solve equations
(\ref{4.7}) and equation (\ref{4.15}) in the limit of zero temperature to
obtain the following results presented in tables 3 and 4.

\begin{center}
\begin{tabular}{|c|c|c|c|}
\hline
$N$ & Ground state strings & Positive energy strings &
Zero energy strings \\
& ($\rho^{(h)}=0$) & ($\rho=0$) & ($\rho=\rho^{(h)}=0$) \\
\hline
even & $\nu-1$ & $\nu$ & the rest \\
\hline
odd & $\nu+2$ & $\nu,\nu+1$ & the rest \\
\hline
\end{tabular}
\end{center}
\begin{center}
Table 3.
\end{center}

\begin{center}
\begin{tabular}{|c|c|c|}
\hline
$N$ & Negative Energy & Positive Energy \\
\hline
even & $\frac{\widehat\epsilon_{\nu-1}(w)}{4p_0I}=-\hat{\rho}_{\nu-1}=
\frac{shw(2s-p_0\frac{3-v_s}{2})}{sh2w}$ &
$\frac{\widehat\epsilon_{\nu}(w)}{4p_0I}=\hat{\rho}_{\nu}^{(h)}=
\frac{shw(2s+2-p_0\frac{3-v_s}{2})}{sh2w}$ \\
\hline
odd & $\frac{\widehat\epsilon_{\nu+2}(w)}{4p_0I}=-\hat{\rho}_{\nu+2}=
2\frac{ch\frac{w}{2}shw(2s-p_0\frac{3-v_s}{2})}{sh2w}$
& $\frac{\widehat\epsilon_{\nu}(w)}{4p_0I}=\hat{\rho}_{\nu}^{(h)}=
\frac{chw(2s+1-p_0\frac{3-v_s}{2})}{chw}$ \\
&  & $\frac{\widehat\epsilon_{\nu+1}(w)}{4p_0I}=\hat{\rho}_{\nu+1}^{(h)}=
2\frac{shw(2s+2-p_0\frac{3-v_s}{2})}
{sh2w}ch\frac{w}{2}$ \\
\hline
\end{tabular}
\end{center}
\begin{center}
Table 4.
\end{center}

\noindent
The symbol "hat" in table 4 stands for the Fourier transform.\\
Before finishing this section, we want to point out that the $\beta\rightarrow
0$ limit of the free energy (\ref{4.13}) provides the justification of the
string hypothesis. Indeed, using results collected in table 2 and the formula
(\ref{4.13}), it is easy to obtain:

\begin{equation}
\lim_{\beta\rightarrow 0}\beta \frac{F}{L}=
-\sum_j|\hat{a_j}(0)|\ln(1+\overline{\eta}_j^{-1})=-\ln N'
\label{4.16}
\end{equation}

\noindent
where $N'=N$ ($N/2$) for $N$ odd (even). Equation above implies
that the total number of states $(N')^L$ is correctly reproduced by the use of
the string hypothesis.

\chapter*{4. The Commensurable N-even Case: N=2 SUSY}

{}From the results of the previous section (tables 3 and 4) we get the relevant
information for unraveling the physics of the $N$ even case. To make the
resulting picture intuitively more clear, we will begin by considering the
commensurable case, point $A$ in figure 2.
For generic $N$ even this
corresponds to the anisotropic antiferromagnet of spin $j=\frac{(N-2)}{4}$ and
anisotropy $\gamma=\frac{2\pi}{N}$.

\setcounter{chapter}{4}
\setcounter{section}{0}
\setcounter{equation}{0}

\section{Commensurable Case: RSOS-structure}

A qualitative picture of the model can be obtained by comparing its string
structure with that of the isotropic model with identical spin. In the
isotropic case we have the same negative energy strings but a different
pattern of the zero energy strings. In fact, all strings with length greater
than $2j$ are allowed and all of them have zero energy. The elementary
excitations for the isotropic model are holes in the Dirac sea of $2j$-strings.
The internal quantum numbers of these elementary excitations are of two types:
a "vertex"-spin with value equal to $1/2$ and an "RSOS"-spin defined by the
Coxeter graph of the type A and Coxeter number $2j+2$. These two types of spin
determine the dimension of the Hilbert space ${\cal H}^{(n)}$ with $n$-holes
\cite{8}:

\begin{equation}
\dim{\cal H}^{(n)}=2^n\sum_{i_1,\dots,i_{n-1}=1}^{2j+1} \overline{K}_{1i_1}
\dots \overline{K}_{i_{n-1}1}
=2^n\sum_{i=1}^{ 2j+1}\frac{\sin^2(\frac{\pi i}{2j+2})}{j+1}(2\cos
(\frac{\pi i}{2j+2}))^n
\label{5.1}
\end{equation}

\noindent
with $\overline{K}$, the incidence $(2j+1)\times(2j+1)$ matrix of the graph
$A_{2j+2}$:

\begin{eqnarray}
\overline{K}= \begin{array}{c}\left|\;\; \begin{array}{lcr} 010 & \dots & 0 \\
101 & \dots & 0 \\
0 & \dots & 101 \\
0 & \dots & 010
\end{array}\;\;
\right | \\
\end{array}
\label{5.2}
\end{eqnarray}

\noindent
The $2^n$ contribution to (\ref{5.1}) can be effectively obtained by ignoring
all the zero energy strings with length smaller than $2j$. In fact, the
counting of the spin $1/2$ degrees of freedom is, essentially, identical to
that of the $XXX$ chain. The RSOS-contribution comes from the zero energy
strings with length smaller than $2j$.

For the commensurable case (see table 4), we do not have zero energy strings
with length greater than $2j$. Instead, we have positive energy negative
parity strings of length 1. By drawing analogy with an isotropic case, we may
expect the same RSOS structure, but somewhat different "vertex" spin
contribution.

{}From the results collected in tables 3 and 4, we conclude (for $N$ even
commensurable case) that the ground state is a Dirac sea of
$(2j=\nu-1)$-strings and
that there exist two types of the elementary excitations, degenerate in energy:
holes of the Dirac sea and particles associated with the negative
parity strings. The particle-hole structure of the commensurable chain is
identical to that of $XX$ model. These qualitative
arguments indicate that the commensurable chain is equivalent to the free
fermions equipped with RSOS-internal degrees of freedom. To prove this claim,
we will proceed in step by step fashion: first, computing the dimension of the
Hilbert space of $n$-elementary excitations, then, deriving scattering $S$-
matrix and, finally, calculating the central extension.

We start defining the spin $S_z$ of the generic Bethe state $|\psi>$ as a
difference in the number of roots with respect to the ground state
\begin{eqnarray}
S_z(|\psi>)=\sharp roots (|\psi>)-\sharp roots (|Gr.\;state>)
\label{5.3}
\end{eqnarray}

\noindent
For $N$ even case we find with a help of the equation (\ref{4.7})
and the expression for the ground state density (table 4) the following:

\begin{equation}
S_z(|\psi>)=\frac{p_0}{2}(\tilde{n}_{\nu}-\tilde{n}_{\nu-1}^{(h)})
\label{5.4}
\end{equation}

\noindent
where $\tilde{n}_{\nu}$ is a number of positive energy strings with negative
parity and $\tilde{n}_{\nu-1}^{(h)}$ is a number of holes in the Dirac ground
state. We can interpret (\ref{5.4}) as a renormalized spin and refer to
$\frac{\tilde{n}_{\nu}-\tilde{n}_{\nu-1}^{(h)}}{2}$ as a spin of the state.
Recalling that $p_0=\frac{N}{2}$ and taking into account that $S_z$
is always an integer number, we obtain the restriction
\begin{eqnarray}
\tilde{n}_{\nu}-\tilde{n}_{\nu-1}^{(h)}= even\;\; integer\;(p_0\;\; odd)
\label{5.5}
\end{eqnarray}

Next, we proceed to compute the dimension of the Hilbert space with
$n$-elementary excitations. The strategy will be to count all degenerate Bethe
states containing the total number of particles ($\nu$-strings) and holes
equal $n$. From the hole-particle degeneration, at the commensurable point, we
readily get $2^n$-states. Additional degeneration will correspond to the zero
energy strings with length smaller than $2j$, i.e. the RSOS part.

To count the RSOS degeneration we should, first, invert the equations
(\ref{4.7}) with a help of the inversion formulas presented in appendix B.
For $N$ even case we derive
\begin{eqnarray}
\rho_1+\rho_1^{(h)} & = & s_1\ast\rho_2^{(h)} \nonumber \\
\rho_j+\rho_j^{(h)} & = & s_1\ast(\rho_{j+1}^{(h)}+\rho_{j-1}^{(h)})\;;\;\;\;
1<j<\nu-2  \nonumber \\
\rho_{\nu-2}+\rho_{\nu-2}^{(h)} & = & s_1\ast(\rho_{\nu-3}^{(h)}+
\rho_{\nu-1}^{(h)}+\rho_{\nu})
\label{5.6} \\
\rho_{\nu-1}+\rho_{\nu-1}^{(h)} & = & s_1\ast\rho_{\nu-2}^{(h)}+
\frac{|\epsilon_{\nu-1}|}{4p_0I} \nonumber \\
\rho_{\nu}+\rho_{\nu}^{(h)} & = & s_1\ast\rho_{\nu-2}^{(h)}+
\frac{\epsilon_{\nu}}{4p_0I} \nonumber
\end{eqnarray}

\noindent
where $s_1$ was defined in section 3. \\
Taking the Fourier transform of (\ref{5.6}) and setting $w=0$, we have for
the non-negative integers
$\tilde{n}_j(\tilde{n}_j^{(h)})=\hat{\rho}_j(w=0)(\hat{\rho}_j^{(h)}(w=0))$
the following system

\begin{eqnarray}
\tilde{n}_1+\tilde{n}_1^{(h)} & = & \frac{1}{2}\tilde{n}_2^{(h)} \nonumber \\
\tilde{n}_j+\tilde{n}_j^{(h)} & = & \frac{1}{2}(\tilde{n}_{j+1}^{(h)}+
\tilde{n}_{j-1}^{(h)})\;;\;\;\; 1<j<\nu-2
\label{5.7} \\
\tilde{n}_{\nu-2}+\tilde{n}_{\nu-2}^{(h)} & = & \frac{1}{2}
\tilde{n}_{\nu-3}^{(h)}+\mbox{\boldmath k} \nonumber
\end{eqnarray}

\noindent
with $\mbox{\boldmath k}=\frac{\tilde{n}_{\nu}+\tilde{n}_{\nu-1}^{(h)}}{2}$.
Eqns. (\ref{5.7}) imply:
\begin{eqnarray}
\mbox{\boldmath k} & = & \sum_{j=1}^{\nu-2}\tilde{n}_j(\nu-1-j)+
\frac{\tilde{n}_1^{(h)}(\nu-1)}{2}
\label{5.8} \\
\tilde{n}_i^{(h)} & = & \sum_{j=1}^{\nu-2}2\tilde{n}_j(i-j)\theta(i-j>0)+
i\tilde{n}_1^{(h)} \nonumber
\end{eqnarray}
\\
\noindent
{}From the eqns. (\ref{5.8}) we get, for a given $\mbox{\boldmath k}$, the set
of pairs $\{\tilde{n}_i,\tilde{n}_i^{(h)}\}$, $1\leq i\leq\nu-2$, which
characterize the RSOS degeneration:

\begin{eqnarray}
\sum\prod_{i=1}^{\nu-2}
\left[ \begin{array}{c} \tilde{n}_i+\tilde{n}_i^{(h)} \\ \tilde{n}_i
\end{array} \right]=\;"count"
\label{5.9}
\end{eqnarray}

\noindent
where

\begin{equation}
"count"=(\overline{K}^{2\mbox{\boldmath{k}}})_{1,1}=\sum_{i=1}^{ 2j+1}
\frac{\sin^2(\frac{\pi i}{2j+2})}{j+1}(2\cos
(\frac{\pi i}{2j+2}))^{2\mbox{\boldmath k}}\;;\;\;\;\;
2\mbox{\boldmath{k}}=\;even\;integer
\label{5.10}
\end{equation}

\begin{equation}
"count"=(\overline{K}^{2\mbox{\boldmath{k}}})_{1,2j+1}=\sum_{i=1}^{ 2j+1}
(-1)^{i+1}\frac{\sin^2(\frac{\pi i}{2j+2})}{j+1}(2\cos
(\frac{\pi i}{2j+2}))^{2\mbox{\boldmath k}}\;;\;\;\;\;
2\mbox{\boldmath{k}}=\;odd\;integer\;,\;\;\mbox{\boldmath{k}}\geq j
\label{5.11}
\end{equation}

\noindent
In the formula (\ref{5.9}) the sum is taken over the whole set of solutions
to (\ref{5.8}).

Even though we don't have an analytical proof of the identity
(\ref{5.9}, \ref{5.10}, \ref{5.11}),
we ran extensive numerical check on it, so that
we are convinced in its validity beyond reasonable doubt.

For $n$-elementary excitations ($n=2\mbox{\boldmath k}$) we get
\begin{eqnarray}
\dim{\cal H}^{(n)}=2^n\times"count"
\label{5.12}
\end{eqnarray}
\noindent
Next, we will consider two possible cases: $p_0$-even integer and $p_0$-odd
integer. The $p_0$-odd integer case corresponds to the spin $j$ integer and the
"count" number (\ref{5.10}) coincides with the multiplicity of the identity
in the tensor product of $2\mbox{\boldmath k}$ copies of the spin $1/2$ irrep,
provided we use the truncated decomposition rule:

\begin{equation}
j_1\times j_2=\sum_{j_1-j_2}^{\min(j_1+j_2,2j-j_1-j_2)}j\;;\;\;\;\;2j=\nu-1
\label{5.13}
\end{equation}
\\
This is a well known decomposition rule for irreps of $\widehat{SU}(2)$
Kac-Moody algebra with level $2j$. The RSOS interpretation is obtained by
associating with each state in ${\cal H}^{(n)}$ a path of length $n+1$ in the
Bratteli diagram defined by the graph $A_{2j+2}$, starting and
finishing at the zero point (see fig. 3).

It is easy to conclude from fig. 3 that if $n$ is odd, then the RSOS
contribution is zero. This fact reflects the restriction $(\tilde{n}_\nu-
\tilde{n}_{\nu-1}^{(h)})=\;even\;integer$ we have mentioned above.

The $p_0$ even integer case corresponds to the spin $j$-half integer. In this
case eqn. (\ref{5.4}) does not impose any restrictions on the number of
particles or holes of the state. For the formula (\ref{5.11}) we obtain
the following Bratteli diagram interpretation:

i) $2\mbox{\boldmath k}=$ even integer: Path in $A_{2j+2}$ start and finish at
zero.

ii) $2\mbox{\boldmath k}=$ odd integer, $\mbox{\boldmath k} \geq j$:
Path in $A_{2j+2}$ start at zero and finish at $j$.

\section{$S$-matrices}

The physics of the model will result from solving the following "inverse
problem", known as higher level Bethe ansatz \cite{30}. Given an arbitrary
Bethe state $|\psi>$, we should look for a scattering $S$-matrix such that
the Bethe equations are equivalent to:

\begin{equation}
e^{ik_j{\overline N}}|\psi>=S(k_j,k_{j+1})\dots S(k_j,k_{\overline N})
\dots S(k_j,k_{j-1})|\psi>
\label{5.14}
\end{equation}

\noindent
The $S$-matrices in (\ref{5.14}) will describe the scattering between the
elementary excitations of the model. From the previous counting, we expect
that the $S$-matrix in (\ref{5.14}) factorizes into the two pieces, each one
associated with one of the two different types of spin.

The string structure of our model at the commensurable point (i.e. presence
of only one string with index greater than $2j$ and particle-hole
degeneracy) can be taken as indication that the spin $1/2$ part of the
$S$-matrix must be the one of the $XX$ model. The following is a simple
argument to make this claim plausible. In the isotropic case, the singlet
and the
triplet states differ by the existence of an extra zero energy string of length
$2j+1$. This extra string is the reason for the difference between the singlet
and triplet $S$-matrices. In the commensurable
case, we do not have this extra string and the "singlet"
(particle-hole) and the "triplet" (particle(hole)-particle(hole),
hole-particle) are practically the same from the point of view of the string
content.

To get the $S$-matrices, we will start with the system of eqns (\ref{5.6})
from which we can obtain:

\begin{equation}
\frac{|\hat\epsilon_{\nu-1}|}{4p_0I}=\hat{\rho}_{\nu-1}+\hat{\rho}_{\nu-1}^{(h)}
-\frac{sh(\nu-2)w}{2chw sh(\nu-1)w}(\hat{\rho}_{\nu}+\hat{\rho}_{\nu-1}^{(h)})
+\frac{sh(\nu-2)w}{sh(\nu-1)w}\hat{\rho}_{\nu-2}+\dots
\label{5.15}
\end{equation}
\\
\noindent
Note that the equation above has the same structure as the original Bethe
ansatz equations (\ref{4.7}), however, with one important distinction: the
roles of $\rho_{\nu}$ and $\rho_{\nu}^{(h)}$ are completely reversed.\\
Let us consider the simplest case: the Bethe state with two holes and one
($\nu-2$)-string with rapidities $u_1$, $u_2$ and $\frac{u_1+u_2}{2}$,
respectively. The hole-hole $S$-matrix will contain two contributions:
one, coming from the scattering between two ($\nu-1$)-holes and the other,
from the scattering between ($\nu-1$)-hole and the ($\nu-2$)-string. These
contributions can be directly read from (\ref{5.15})

\begin{eqnarray}
\ln S_{\nu-1,\nu-1}(u) & = & -\frac{i}{2}\int_{-\infty}^{\infty}
\frac{dw}{w}\sin wu\frac{sh(\nu-2)w}{chwsh(\nu-1)w} \\
\label{5.16}
\partial_u\ln S_{\nu-1,\nu-2}(u) & = & i\int_{-\infty}^{\infty}
dw\cos wu\frac{sh(\nu-2)w}{sh(\nu-1)w} \nonumber
\end{eqnarray}

\noindent
The hole-hole $S$-matrix is given by:
\begin{equation}
S(u_{21})=S_{\nu-1,\nu-1}(u_{21})S_{\nu-1,\nu-2}(\frac{u_{21}}{2})=
e^{-i\int_{-\infty}^{\infty}\frac{dw}{w}\sin\frac{u_{21}w}{2}
\frac{sh\frac{\nu-2}{2}w}{ch\frac{w}{2}sh\frac{\nu-1}{2}w}}
\times\frac{sh\frac{\pi}{2(\nu-1)}(\frac{u_{21}}{2}-i)}{sh\frac{\pi}{2(\nu-1)}
(\frac{u_{21}}{2}+i)}
\label{5.17}
\end{equation}
\\
Using the same technique, we can easily check that the particle-hole and the
particle-particle $S$-matrices are identical to (\ref{5.17}). This implies
that the particle-hole system is, indeed, behaving as $XX$-chain with respect
to the vertex spin. The formula (\ref{5.17}) gives us the RSOS-contribution
to the scattering $S$-matrix. For the case of the two elementary excitations
the RSOS contribution is rather simple, in fact, it corresponds to the
RSOS-Boltzman weight

$$
\left[ \begin{array}{cc} \frac{1}{2} & id \\ id & \frac{1}{2} \end{array}
\right](u)
\nonumber
$$

\noindent
for the $A_{2j+2}$-model.

Thus, we can represent our $S$-matrices in the following symbolic form:

\begin{equation}
S_{XX}\bigotimes S_{RSOS(p=1,r=2j+2)}
\label{5.18}
\end{equation}

\noindent
Notice, that the $S$-matrix we got, has the same structure as the $S$-matrix
for $N=2$ integrable models \cite{16}:

\begin{equation}
S^{N=2}=S_{A_2}^{N=2}\bigotimes S_{A_{2j+2}}^{N=0}
\label{5.19}
\end{equation}
\\
\noindent
where $S_{A_2}^{N=2}$ is the Sine-Gordon $S$-matrix for $\beta^2=8\pi 3/2$
(which corresponds to $q=1$ with $q=e^{i\pi /p}$ and $p=\frac{\beta^2}
{8\pi-\beta^2}$). The $N=2$ SUSY is not explicit in the Sine-Gordon model,
but is realized through the quantum group symmetry in the nonlocal way
\cite{12}, \cite{14}.

\section{Central Extensions}

Our analysis of the $S$-matrices together with the counting of states led us to
the conclusion that the commensurable chain is equivalent to the tensor
product of an $XX$ spin $1/2$-chain and an RSOS model based on the graph
$A_{2j+2}$. This picture will be consistent if the central charge is given by:
\begin{equation}
c=1+(2-\frac{6}{2j+2})=\frac{3j}{j+1}\;;\;\;\;\;2j+1=\frac{N}{2}
\label{5.20}
\end{equation}

\noindent
Two pieces which appear in the sum above are due to the $XX$ and RSOS
contributions, respectively.

To see that this is, indeed, the case we now proceed to calculate the central
extension from the low temperature expansion for the entropy. Following along
the standard lines \cite{9}, we differentiate the system (\ref{4.15}) with
respect to $u$ and compare the result with the system (\ref{5.6}). Then, it is
trivial to establish asymptotic validity of the following formula:

\begin{equation}
\rho_j\approx\frac{1}{2\pi v_{sound}}\frac{(-1)^{r_j}\partial_u \epsilon_j}
{1+\eta_j}\;;\;\;\;\;v_{sound}=p_0I\;;\;\;u\gg 1
\label{5.21}
\end{equation}
\\
\noindent
Plugging (\ref{5.21}) into the expression for the entropy $S$

\begin{eqnarray}
\frac{S}{L} & = & \sum_j \int du [(\rho_j+\rho_j^{(h)})
\ln(\rho_j+\rho_j^{(h)})-\rho_j\ln\rho_j-\rho_j^{(h)}\ln\rho_j^{(h)}]
\nonumber \\
& = & \sum_j \int du\rho_j(u)[(1+\eta_j)\ln(1+\eta_j)-\eta_j\ln\eta_j]
\label{5.22}
\end{eqnarray}

\noindent
we obtain

\begin{equation}
S\approx 2TL\sum_{j=1}^{\nu}\frac{(-1)^{r_j}}{\pi v_{sound}}
\{ {\cal L}(\frac{1}{1+\overline{\overline{\eta}}_j})-
{\cal L}(\frac{1}{1+\overline{\eta}_j})\}
\label{5.23}
\end{equation}
\\
\noindent
where $\cal L$ stands for the dilogarithmic Roger's function, defined as

\begin{equation}
{\cal L}(x)=-\frac{1}{2}\int_0^x dx\{\frac{\ln x}{1-x}+\frac{\ln(1-x)}{x}\}
\label{5.24}
\end{equation}

\noindent
Taking $\overline{\eta}_j,\overline{\overline{\eta}}_j$ from table 1 and
exploiting "magic" formulas for the dilogarithmic functions \cite{31}:

\begin{equation}
\left\{ \begin{array}{lll}
\sum_{k=2}^{n-2}{\cal L}(\frac{\sin^2\frac{\pi}{n}}{\sin^2\frac{\pi k}{n}})
& = & (1-\frac{3}{n})\frac{\pi^2}{3}  \\
\sum_{j=2}^n{\cal L}(\frac{1}{j^2}) & = & \frac{\pi^2}{6}-2{\cal L}
(\frac{1}{n+1}) \\
{\cal L}(x)+{\cal L}(1-x) & = & \frac{\pi^2}{6}
\end{array} \right.
\label{5.25}
\end{equation}

\noindent
we, finally, obtain for $S$ and the central extension $c$ the following:

\begin{equation}
S\approx\frac{2\pi TL}{v_{sound}}(\frac{1}{2}-\frac{1}{\nu+1})
\label{5.26}
\end{equation}

\begin{equation}
c=\frac{3v_{sound}}{\pi TL}\frac{\partial S}{\partial T}
=\frac{3(\frac{\nu-1}{2})}{\frac{\nu-1}{2}+1}
\label{5.27}
\end{equation}

\noindent
Clearly, the formula (\ref{5.27}) above is in perfect agreement with our prior
anticipation (\ref{5.20}).

\section{Integrable Deformations}

The $XX$-part of our model, by itself, admits nilpotent extension as we have
discussed in section 2. This extension corresponds to an addition of an extra
magnetic field coupled only to spin $1/2$ degrees of freedom. The product
of the nilpotent $R$-matrix ($N=4$) and an RSOS $S$-matrix

\begin{equation}
R^{\lambda\lambda}_{N=4}\bigotimes S_{RSOS(p=1,r=2j+2)}
\label{5.28}
\end{equation}

\noindent
describes our model away from the commensurable point (i.e. for generic spin).
To check this claim, we, first, observe that the energies of holes
$\epsilon_{\nu-1}$ and particles $\epsilon_{\nu}$ presented in table 4
are, essentially, equivalent to those of the nilpotent $XX$-chain. Second, the
RSOS counting formula (\ref{5.9}) remains valid and so does our central charge
result (\ref{5.27}). Note, however, that the particle-hole degeneracy is now
irretrievably lost.

We now explain the somewhat surprising result that the central extension
does not show any dependence on our nilpotent magnetic field related to
the irregular spin $s$. The point is that this internal magnetic field is
selective in nature: it affects only fermionic degrees of freedom, but not
RSOS degrees of freedom. Since, the central charge for the free fermions
(with or without magnetic field) is equal to 1 and an RSOS part is immune
to the internal magnetic field, we conclude that all conformal degrees of
freedom are preserved and, therefore, central charge does not change when we
move away from the commensurable point. This is contrary to the behaviour of
the higher spin $XXZ$ model in the external magnetic field, which freezes out
all parafermionic degrees of freedom and reduces the central charge from
$WZW$-value to just 1.

We conclude this section with the following observation. We started by
introducing the model based on the nilpotent representation of quantum group.
This nilpotency somehow managed to survive the "hell" of the
renormalization (filling in Dirac sea, etc.) and reemerged (like Phoenix) in
the
modified form in the formula for the physical $S$-matrices, as expression
(\ref{5.28}) indicates.

\chapter*{5. $N$ Odd Case: Beyond Bootstrap}

For $N$ odd, the commensurable case corresponds to ($\gamma=\frac{2\pi}{N},
j=\frac{N-1}{2}$). Results collected in tables 3 and 4 of section 3,
clearly indicate that the string structure of the $N$ odd case is a bit more
involved than that of the $N$ even case, and the departure from the
properties of an isotropic model ($\gamma=0,j=\frac{N-1}{2}$) is also more
dramatic. First of all, the ground state strings are of length $2j_{eff}$
with $j_{eff}=\frac{j}{2}$ and not, as it could be naively expected, of length
$2j$. This circumstance should be interpreted as an indication that for an
$N$ odd case our model is in the strong anisotropic regime.
The strings with length smaller than $2j_{eff}$ are of the zero energy
which indicates the presence of an RSOS structure based on the
$A_{2j_{eff}+2}$-graph. The pattern
of the positive energy strings is also new. For the $N$ even case, we have only
the negative parity strings degenerate in energy with holes of the Dirac sea.
For the $N$ odd case, we have two types of positive energy strings: one with
length $2j+3$ and another, negative parity string of length $1$, designated
as ($\nu+1$) and $\nu$-strings in tables 3 and 4. At the commensurable point,
the ($\nu+1$)-string and the holes of the Dirac sea are degenerate in energy.
In this section we will describe some aspects of the system defined by the
holes of the Dirac sea and the two positive energy strings in terms of the
RSOS model with $p=2$ and the restriction parameter $r=6$ \cite{22}.
The emerging picture
is drastically different from the one we have found in the $N$ even case where
the model was shown to be equivalent to the tensor product of a vertex $XX$
model and an RSOS model based on the $A_{2j+2}$ graph. Notice that the
($p=2, r=6$) RSOS model which, for $N$ odd, replaces the $XX$-part of the
$N$ even case, describes the spin $1$, $SU(2)_q$-invariant
Fateev-Zamolodchikov chain with $q^6=1$ \cite{32}.

\setcounter{chapter}{5}
\setcounter{section}{0}
\setcounter{equation}{0}

\section{Counting of States}

The first thing we would like to do is to count the degeneration of the Bethe
states following, essentially, the same procedure we used for the $N$ even
case. Once again we invert eqns. (\ref{4.7}) with a help of the identities
presented in appendix B to get

\begin{eqnarray}
\rho_j+\rho_j^{(h)} & = & s_1\ast [\rho_{j+1}^{(h)}+\rho_{j-1}^{(h)}]+
\delta_{j,\nu-1}[s_1\ast\rho_j^{(h)}- \nonumber \\
& - & (s_2+s_1)\ast\rho_{\nu}^{(h)}+f(u)] \;;\;\;\;\;1\leq j\leq\nu-1
\label{6.1} \\
\rho_{\nu}+\rho_{\nu}^{(h)} & = & s_2\ast[\rho_{\nu+1}^{(h)}+\rho_{\nu-1}^{(h)}
+\rho_{\nu+2}] \nonumber \\
\rho_{\nu+1+\tilde{r}}+\rho_{\nu+1+\tilde{r}}^{(h)}
& = & s_2\ast \rho_{\nu}^{(h)}+\frac{(-1)^{\tilde{r}}}{2}
\frac{\cos 2\pi x}{ch\pi u+(-1)^{\tilde{r}}\sin 2\pi x}\;;\;\;\;
\tilde{r}\equiv 0,1 \nonumber
\end{eqnarray}
\\
\noindent
Here, notations are the same as in section 3. Applying the Fourier transform to
the system above and setting $w=0$, we derive, after a bit of labor, the
following:

\begin{eqnarray}
\tilde{n}_1+\tilde{n}_1^{(h)} & = & \frac{1}{2}\tilde{n}_2^{(h)} \nonumber \\
\tilde{n}_j+\tilde{n}_j^{(h)} & = & \frac{1}{2}(\tilde{n}_{j+1}^{(h)}+
\tilde{n}_{j-1}^{(h)})\;;\;\;\; 1<j\leq\nu-2
\label{6.2} \\
\tilde{n}_{\nu-1}+\tilde{n}_{\nu-1}^{(h)} & = & \frac{1}{2}
\tilde{n}_{\nu-2}^{(h)}+\tilde{\mbox{\boldmath k}} \nonumber
\end{eqnarray}

\noindent
where
\begin{equation}
\tilde{\mbox{\boldmath k}}=\tilde{n}_{\nu}+\frac{\tilde{n}_{\nu+1}+
\tilde{n}_{\nu+2}^{(h)}}{2}\;;\;\;\;\;2\tilde{\mbox{\boldmath k}}=
\;even\;integer
\label{6.3}
\end{equation}
\\
\noindent
Comparing system (\ref{6.2}) with system (\ref{5.7}), one cannot help
noticing that they are remarkably similar. In fact, the relation between
them is simply
\begin{eqnarray}
\mbox{\boldmath k} & \rightarrow & \tilde{\mbox{\boldmath k}} \\
\label{6.4}
\nu & \rightarrow & \nu+1 \nonumber
\end{eqnarray}

\noindent
Note, however, that unlike $2\mbox{\boldmath k}$, $2\tilde{\mbox{\boldmath k}}$
is always an even integer number.

Repeating the same analysis as in section 4, it is straightforward to infer
that the total number of the degenerate Bethe states with the same
$\{\tilde{\mbox{\boldmath k}},\tilde{n}_{\nu}\}$ is a product of two factors.
The first RSOS factor is due the to zero energy strings contribution and is
equal to the number of paths of length $2\tilde{\mbox{\boldmath k}}$
in the $A_{2j_{eff}+2}$-Bratteli diagram, which start and end at zero point.
The second factor $2^{2(\tilde{\mbox{\boldmath k}}-\tilde{n}_{\nu})}$ is
related to the degeneracy between holes and $(\nu+1)$-strings. Note that this
degeneracy disappears once we move away from the commensurable point and,
consequently, the second factor reduces to 1. \\
Again, we can describe RSOS factor as the multiplicity of the identity in the
tensor product of
$2\tilde{\mbox{\boldmath k}}$-copies of the $1/2$-irrep. This tensor product is
defined by the truncated decomposition rule:

\begin{equation}
j_1\times j_2=\sum_{j_1-j_2}^{\min(j_1+j_2,2j_{eff}-j_1-j_2)}j
\label{6.5}
\end{equation}
\\
\noindent
The explicit formula for the RSOS factor ($\tilde{\mbox{\boldmath k}}$-fixed)
is the same as (\ref{5.10}) with $j_{eff}$ substituted for $j$ and
$\tilde{\mbox{\boldmath k}}$ is substituted for $\mbox{\boldmath k}$.

For the spin $S_z$ of the Bethe state $|\psi>=|\tilde{n}_{\nu},
\tilde{n}_{\nu+1},\tilde{n}_{\nu+2}^{(h)}>$ one finds

\begin{eqnarray}
S_z(|\psi>)=\sharp roots (|\psi>)-\sharp roots (|Gr.\;state>)=\frac{N}{2}
(\tilde{n}_{\nu+1}-\tilde{n}_{\nu+2}^{(h)})
\label{6.6}
\end{eqnarray}

\noindent
This implies (for $N$ odd) the restriction
\begin{eqnarray}
(\tilde{n}_{\nu+1}-\tilde{n}_{\nu+2}^{(h)})=\;even\;integer
\label{6.7}
\end{eqnarray}

\noindent
{}From (\ref{6.6}) we also observe that the negative parity strings do not
contribute to the total spin of the state, which indicates that the
$\nu$-string behaves as a spin "singlet".

Ignoring, for the time being, the zero energy strings, we now proceed to map
the Bethe states
$|\tilde{n}_{\nu},\tilde{n}_{\nu+1},\tilde{n}_{\nu+2}^{(h)}>$ onto the states
of the $(p=2,r=6)$ RSOS model. This model is defined by the disconnected graph
(see fig. 4). We will limit ourselves to the states defined by the first
graph I. In terms of this graph we can identify the holes and $\nu+1$-strings
with the 2-point links ($j_1,j_2$); $|j_1-j_2|=2$ and the $\nu$-string with
the closed link ($3,3$). We will also impose on the RSOS states periodic
boundary conditions. It is easy to check that this identification is
compatible with the counting and restriction (\ref{6.7}) we have described
above.
In fact, for the given values $\tilde{n}_{\nu},\tilde{n}_{\nu+1}$ and
$\tilde{n}_{\nu+2}^{(h)}$, the corresponding number of the RSOS states related
to graph I is $2^{(\tilde{n}_{\nu+1}+\tilde{n}_{\nu+2}^{(h)})}$. In this way
we capture the "singlet" nature of the $\nu$-particle. Restriction (\ref{6.7})
can
be easily translated into the RSOS language as a requirement that one should
consider only RSOS states which start and finish at the same point on the
graph. If we use graph I to describe holes and positive energy strings then
we will find, as a basis of the Hilbert space with the total number of
$n$-elementary excitations, the following:

\begin{equation}
(j_1,j_2,\dots,j_{n+1})\bigotimes(J_1,\dots,J_{N+1})
\label{6.8}
\end{equation}

\noindent
where
\begin{eqnarray}
\tilde{n}=\tilde{n}_{\nu}+\tilde{n}_{\nu+1}+\tilde{n}_{\nu+2}^{(h)}\;
& \;\;\;\;j_1=j_{n+1}\; & \;\;\;\;j\in\{1,3,5\} \\
\label{6.9}
N=2\tilde{n}_{\nu}+\tilde{n}_{\nu+1}+\tilde{n}_{\nu+2}^{(h)}\;
& \;\;\;\;J_1=J_{N+1}=0\; & \;\;\;\;J\in\{0,\frac{1}{2},\dots,j_{eff}\}
\nonumber
\end{eqnarray}
\noindent
and
\begin{eqnarray}
|J_i-J_{i+1}| & = & \frac{1}{2} \\
\label{6.10}
|j_i-j_{i+1}| & = & \left\{ \begin{array}{ll}
2 & for\;\;(\tilde{n}_{\nu+1}+\tilde{n}_{\nu+2}^{(h)})\;couples \\
0 & and\;\;j_i=3\;for\;\tilde{n}_{\nu}\;couples \\
\end{array} \right. \nonumber
\end{eqnarray}

\noindent
States (\ref{6.8}) for the different partitions of $n$ into the sum of the
three positive integers provide a basis of the Hilbert space with
$n$-elementary excitations.

The basis (\ref{6.8}), which is already compatible with our counting, implies
that
the $S$-matrix for the elementary excitations is a product of the two RSOS
$S$-matrices. In the next section, we will show that this is, in fact,
the case. Before passing to this issue, let us briefly comment on graph II.
This graph can be interpreted as describing strings in a discrete target
space characterized by the graph $A_3$ \cite{33}. According to this graph,
we can
suggest the following "stringy" interpretation. The holes and $(\nu+1)$-strings
can be associated with "closed" RSOS configurations ($2,2$) and ($4,4$),
and the negative parity string $\nu$ with "open" RSOS configuration ($2,4$).
If we maintain this "stringy" interpretation of the graph II, holes
and $(\nu+1)$-strings will appear as bound states of $\nu$-strings.

\section{S-matrices and Bootstrap}

Let us consider the ($p=2,r=6$) RSOS model which is a descendant of the
($p=1,r=6$) model and is obtained by the standard fusion procedure \cite{34}.
This model can live in two different regimes:

\begin{tabbing}
regime (I): \=                                                            \kill
regime (I): \> The ground state is a Dirac sea of length $p=2$-strings and \\
            \> there are no positive energy strings. The central extension \\
            \> is given by:
\end{tabbing}
$$
c=\frac{3p}{p+2}(1-\frac{2(p+2)}{r(r-p)})=1\;;\;\;\;\;(p=2,r=6)
$$
\begin{tabbing}
regime (II): \=                                                           \kill
regime (II): \> The ground state is a Dirac sea of length $r-2=4$-strings. \\
             \> Strings with length $1,2,3=(r-3)$ are of the positive energy.\\
             \> Holes are not allowed in the $(r-2)$ Dirac sea. \\
             \> The central extension is given by:
\end{tabbing}
$$
c=2-\frac{6}{r}=1\;;\;\;\;\;r=6
$$
\\
\noindent
The elementary excitations in regime (I) are holes of the $p=2$ Dirac sea,
while in regime (II) they are associated with the positive energy strings.
Notice that for ($p=2,r=6$) the central extension is the same in both regimes.
To match the dynamics of ($\nu+1,\nu$)-strings and holes of our model
we will choose regime (II).

The Bethe equations of the ($p=2,r=6$) model can be written as follows
\cite{22}:

\begin{equation}
a_{j,2}^{(4)}=\rho_j^{(h)}+\sum_{k=1}^3 A_{j,k}^{(4)}\ast\rho_k \;;
1\leq j\leq 3
\label{6.11}
\end{equation}

\noindent
with the functions $a_{j,2}^{(4)}$ and $A_{j,k}^{(4)}$ best defined by their
Fourier transforms:

\begin{eqnarray}
\hat{a}_{j,2}^{(4)}(w) & = & \frac{1}{2ch\frac{w}{2}} \hat{A}_{j,2}^{(4)}(w)
\nonumber \\
\hat{A}_{j,k}^{(4)}(w) & = & 2\hat{a}_j(w)\hat{n}_k(w)\;,\;\;\;\;j\geq k
\label{6.12} \\
\hat{a}_j(w) & = & \frac{sh(\frac{4-j}{2}w)}{sh(2w)}\;;\;\;\;\;
\hat{n}_k(w)=cth(\frac{w}{2})sh(\frac{kw}{2}) \nonumber
\end{eqnarray}

\noindent
To compute $\hat{A}_{j,k}^{(4)}$ for $j<k$ we should use the symmetries of the
model \cite{22}:

\begin{eqnarray}
\hat{A}_{1,2}^{(4)} & = & \hat{A}_{3,2}^{(4)} \nonumber \\
\hat{A}_{2,3}^{(4)} & = & \hat{A}_{2,1}^{(4)}
\label{6.13} \\
\hat{A}_{1,3}^{(4)} & = & \hat{A}_{3,1}^{(4)} \nonumber
\end{eqnarray}

\noindent
In the regime (II), the positive energies of $1,2,3$-strings are given by:
\begin{equation}
\epsilon_j=a_{j,2}^{(4)}
\label{6.14}
\end{equation}

\noindent
Using (\ref{6.12}) and (\ref{6.13}) we conclude that the 1 and 3 strings
are degenerate in energy
\begin{eqnarray}
\hat{\epsilon}_1=\hat{\epsilon}_3 & = & \frac{1}{2chw}
\label{6.15} \\
\hat{\epsilon}_2 & = & \frac{ch(\frac{w}{2})}{chw} \nonumber
\end{eqnarray}

\noindent
This will agree with our counting if we identify the 1 and 3 strings of the
($p=2,r=6$) RSOS model with the holes and $\nu+1$-strings. However, the
energies do not match in a peculiar way. From table 4, we get (up to the
overall factor)

\begin{eqnarray}
\hat{\epsilon}_{\nu+2} & = & \hat{\epsilon}_{\nu+1}=\frac{ch(\frac{w}{2})}{chw}
\label{6.16} \\
\hat{\epsilon}_{\nu} & = & \frac{1}{chw} \nonumber
\end{eqnarray}

\noindent
These are the values for the commensurable point for arbitrary $N$ odd.
Comparing (\ref{6.15}) and (\ref{6.16}) we, first, observe that for the
($p=2,r=6$) RSOS model the string 2 may be interpreted as a bound states
of two 1-strings. The $\nu+2$-holes and $\nu+1$-strings
have the same energy as 2-string of an RSOS model, however, the negative
parity $\nu$-string has twice the energy of the 1, 3 strings. Furthermore, it
is straightforward to show that this mismatch of energies persists through the
whole hermiticity interval (\ref{3.38}). Before we give an interpretation of
these results, let us see how $S$-matrices, or better to say, the scattering
shifts of the $N$ odd case and ($p=2,r=6$) RSOS model fit together.

The scattering shifts for the interchange of $j$ and $k$-strings are
related to the kernels $A_{j,k}^{(4)}$ in (\ref{6.11}). The $(p=2,r=6$) RSOS
model
is a descendant of the ($p=1,r=6$) RSOS model which completely determines the
Bethe strings scattering. The explicit dependence on the value of $p=2$
only appears in the energies (\ref{6.15}) i.e. in the left-hand side of the
Bethe eqns (\ref{6.11}). Using (\ref{6.12}, \ref{6.13}) we find

\begin{eqnarray}
\hat{A}_{1,1}^{(4)}-1\equiv \hat{A}_{3,3}^{(4)}-1 & = &
\hat{A}_{1,3}^{(4)}\equiv \hat{A}_{3,1}^{(4)}=\frac{1}{2chw}
\;\;\;\;\;\;\;\;\;\;\;\;\;\;\;\;\;\;\;\;\;(a)
\nonumber \\
\hat{A}_{2,2}^{(4)} & = & 1+\frac{1}{chw}
\;\;\;\;\;\;\;\;\;\;\;\;\;\;\;\;\;\;\;\;\;\;\;\;\;\;\;\;\;\;\;\;\;\;\;\;\;\;(b)
\\
\label{6.17}
\hat{A}_{2,3}^{(4)}\equiv \hat{A}_{2,1}^{(4)} & = &
\hat{A}_{3,2}^{(4)}\equiv \hat{A}_{1,2}^{(4)}=\frac{ch\frac{w}{2}}{chw}
\;\;\;\;\;\;\;\;\;\;\;\;\;\;\;\;\;\;\;\;\;\;\;(c) \nonumber
\end{eqnarray}

\noindent
It is important to notice the existence of the extra symmetry reflected in
eqns. (a, c) above which is peculiar for the $(p=2,r=6)$ model. Using the
results above it is trivial to obtain for the $S$-matrices the following:

\begin{equation}
\left\{ \begin{array}{rcl}
S_{1(3),1(3)} & = & tan(\frac{\pi}{4}+i\frac{\pi u}{4}) \\
\label{6.18}
S_{2,2} & = & tan^2(\frac{\pi}{4}+i\frac{\pi u}{4}) \nonumber
\end{array} \right.
\end{equation}
\noindent
Rewriting the Bethe ansatz eqns. (\ref{4.7}) in such a way that roles of
$\rho_{\nu+2}$ and $\rho_{\nu+2}^{(h)}$ are reversed, we get

\begin{eqnarray}
\frac{\hat\epsilon_\nu}{4p_0I} & = & \hat\rho_\nu+\hat\rho_\nu^{(h)}+
\frac{ch(\frac{w}{2})}{chw} (1-\frac{sh(\nu-1)w}{sh\nu w})
(\hat\rho_{\nu+2}^{(h)}+\hat\rho_{\nu+1})+ \nonumber \\
& + & (\frac{1}{chw} -\frac{sh(\nu-1)w}{sh\nu w} -
\frac{sh(\nu-1)w}{chwsh\nu w}) \hat\rho_\nu+ \nonumber \\
& + & 2ch(\frac{w}{2}) \frac{sh(\nu-1)w}{sh\nu w} \hat\rho_{\nu-1} +
2ch(\frac{w}{2}) \frac{sh(\nu-2)w}{sh\nu w} \hat\rho_{\nu-2}+\dots
\label{6.19} \\
-\frac{\hat\epsilon_{\nu+2}}{4p_0I} & = & \hat\rho_{\nu+2}
+\hat\rho_{\nu+2}^{(h)}+ (\frac{1}{2chw}-\frac{sh(\nu-1)w}{2chwsh\nu w})
(\hat\rho_{\nu+2}^{(h)}+\hat\rho_{\nu+1})+ \nonumber \\
& + & \frac{ch(\frac{w}{2})}{chw} (1-\frac{sh(\nu-1)w}{sh\nu w})\hat\rho_\nu+
\frac{sh(\nu-1)w}{sh\nu w}\hat\rho_{\nu-1}+\dots \nonumber
\end{eqnarray}
\noindent
Reading off the scattering shifts from the r.h.s. of (\ref{6.19}), we have

\begin{eqnarray}
S_{hole,hole}(u_{21}) & = &
e^{-i\int_{-\infty}^{\infty}\frac{dw}{w}\sin(u_{21}w)
\frac{sh(\nu-1)w}{chw sh\nu w}}
\times\frac{sh\frac{\pi}{2\nu}(\frac{u_{21}}{2}-i)}{sh\frac{\pi}{2\nu}
(\frac{u_{21}}{2}+i)}
\times tan(\frac{\pi}{4}+i\frac{\pi u_{21}}{4}) \nonumber \\
\label{6.20} \\
S_{\nu,\nu}(u_{21}) & = &
e^{-2i\int_{-\infty}^{\infty}\frac{dw}{w}\sin(u_{21}w)
\frac{sh(\nu-1)w}{chw sh\nu w}} \nonumber \\
& \times & \frac{sh\frac{\pi}{2\nu}(\frac{u_{21}}{2}-i)}{sh\frac{\pi}{2\nu}
(\frac{u_{21}}{2}+i)}
\frac{sh\frac{\pi}{2\nu}(\frac{u_{21}}{2}-\frac{3}{2}i)}
{sh\frac{\pi}{2\nu}(\frac{u_{21}}{2}+\frac{3}{2}i)}
\frac{sh\frac{\pi}{2\nu}(\frac{u_{21}}{2}-\frac{5}{2}i)}
{sh\frac{\pi}{2\nu}(\frac{u_{21}}{2}+\frac{5}{2}i)}
\times tan^2(\frac{\pi}{4}+i\frac{\pi u_{21}}{4}) \nonumber
\end{eqnarray}

\noindent
Here, the first expression describes an evolution of the state consisting of
two ($\nu+2$)-holes and one ($\nu-1$)-string with the rapidities
$u_1,u_2$ and $\frac{u_1+u_2}{2}$, respectively. The second one corresponds to
the state which has two $\nu$-particles with rapidities ($u_1,u_2$) and one
($\nu-2$)-string with the rapidity $\frac{u_1+u_2}{2}$.
Let us summarize the salient features of the equations above. They, clearly,
indicate that our $S$-matrix is a product of two distinctly different
$S$-matrices. The first one is related to the RSOS defined
by graph $A_{2j_{eff}+2}$. The second one is due to ($p=2,r=6$) RSOS model.
Note that the second RSOS comes without the
usual "vacuum-polarization" contribution. Also note that $\nu$-string behaves
as a bound state of two holes. This is not consistent with our previous result
(\ref{6.16}), thus signaling the breakdown of the bootstrap procedure.

Once we have shown the matching of the scattering shifts between the $N$ odd
model and the product of the two RSOS models:
$(p=1,r=2j_{eff}+2)\bigotimes(p=2,r=6)$ we can return to the question of the
energies mismatch we have left open.
The reader should be aware that the Bethe study of the scattering shifts
we are performing here, is not sufficient to uniquely determine the dynamics
of the holes, $\nu+1$ and $\nu$ strings. In fact, the scattering shifts
will be the same for ($p=1, r=6$) and for ($p=2, r=6$) RSOS models. Our
strategy is to get pieces of the puzzle. First, we observe that the RSOS model
with the restriction parameter $r=6$ correctly captures the scattering shifts.
Second, use of ($p=2, r=6$) RSOS-descendant model leads to the correct counting
of the Bethe states, gives the right central extension, however, fails
to reproduce energies of the three types of strings which define the
spectrum of the $N$ odd case. Two comments are now in order.
First of all, a mismatch between bootstrap properties of energies
and of scattering shifts
is typical of the nonfusion models and, therefore, we can imagine that the
model we are seeking is some nonfusion descendant of the ($p=2, r=6$) RSOS
model. Secondly and perhaps more likely, we can attempt to make contact with
the $Z_3$ parafermionic extension of the $N=2$ SUSY $S$-matrices defined in
\cite{16}, i.e. to match our spectrum with the one of the affine Toda
$SU(3)$ (also see comment at the end of this section). If this is the case,
we arrive at the nice physical picture that the
commensurable spin chains for the $N$ even and $N$ odd cases provide a useful
framework where to study the $Z(2)$ and $Z(3)$ $N=2$ integrable models.
Finally, we would like to draw the following important comparison. The most
fundamental difference between $N$ even and $N$ odd cases is that for the
$N$ even, the commensurable point sits just on the border of the weak
anisotropic regime, while for $N$ odd, the commensurable point lies in the
strong anisotropic regime. The analogous difference exists between the massive
$N=2$ model perturbed by the least relevant operator and $N=2$ model perturbed
by the most relevant operator. To complete this section we move to the
computation of the central extension.

Taking advantage of the relevant results from table 2, formulas
(\ref{5.23}, \ref{5.25}) and additional identity

\begin{equation}
{\cal L}(\frac{1}{(1+\nu^{-1})^2})=2{\cal L}(\frac{\nu}{\nu+1})-
2{\cal L}(\frac{\nu}{2\nu+1})
\label{6.21}
\end{equation}

\noindent
we have for entropy $S$

\begin{eqnarray}
S & = & \frac{2TL}{\pi v_{sound}}
\{ \sum_{k=2}^{\nu}[{\cal L}(\frac{\sin^2\frac{\pi}{\nu+2}}
{\sin^2\frac{\pi k}{\nu+2}})-
{\cal L}(\frac{1}{k^2})]+{\cal L}(1)+
{\cal L}(\frac{1}{(1+\nu^{-1})^2})+ \nonumber \\
& + & {\cal L}(\frac{\nu}{2\nu+1})-{\cal L}(\frac{1}{2-(\nu+1)^{-1}})\}
=\frac{\pi TL}{v_{sound}}(1-\frac{4}{N+3})
\label{6.22}
\end{eqnarray}

\noindent
and for the central charge

\begin{equation}
c=\frac{3v_{sound}}{\pi TL}\frac{\partial S}{\partial T}
=\frac{3j_{eff}}{j_{eff}+1}\;;\;\;\;\;j_{eff}=\frac{N-1}{4}
\label{6.23}
\end{equation}
\\
\noindent
The RSOS representation of our model

\begin{equation}
RSOS_{p=2,r=6}\bigotimes RSOS_{p=1,r=2j_{eff}+2}
\label{6.24}
\end{equation}

\noindent
gives us the central extension

\begin{equation}
c=(2-\frac{6}{6})+(2-\frac{6}{2j_{eff}+2})=\frac{3j_{eff}}{j_{eff}+1}\;,
\label{6.25}
\end{equation}
\\
\noindent
provided that both RSOS models are in the regime (II). \newline
Clearly, the formula above is in perfect agreement with our exact result
(\ref{6.23}).

We would like to finish this section in somewhat speculative vein to express
what may be the most natural extension of $N$ even case results to $N$ odd
case (or, more generally, to generic $q$-root of unity case).
Since $N$ even
case describes free fermions "dressed" by RSOS degrees of freedom, then it
is plausible that the generic $q$-root of unity case corresponds to free
parafermions upgraded by RSOS degrees of freedom. These free parafermions
are "free" in a sense of possessing additional symmetry related to the Onsager
algebra (see ref. \cite{35} for the latest review).
We hope to expand on these matters in our future publications.

\chapter*{6. Magnetic Properties of Nilpotent Spin Chain}

In this section, we briefly discuss behaviour of our nilpotent spin chains in
the external magnetic field ($h$). In ref. \cite{21} we have shown that at
$h=T=0$ our system exhibits phenomenon of ferrimagnetism, that is, the ground
state spin is greater than zero-antiferromagnetic value, but smaller than
maximum possible
ferromagnetic value. From the discussion given in the previous sections, it is
abundantly clear that this nonvanishing magnetization is due to the presence
of "internal magnetic field", which selectively effects available degrees of
freedom. By placing the system in the external magnetic field, one may hope to
demonstrate intricate interplay between these two fields, and to "probe" the
model beyond extreme long distance universality, which is, to extend,
nilpotency "blind".

Let us choose "up" small magnetic field in such a way that the ground state
Fermi band shrinks from $-\infty\leq u\leq\infty$ to $-B\leq u\leq B$ where
$B\gg 1$ is the Fermi momentum. Following Yang and Yang \cite{36}, we have
for $T=0$

\setcounter{chapter}{6}
\setcounter{section}{0}
\setcounter{equation}{0}

\begin{equation}
\rho_{j_0}(u+B)=\rho_+(u)+\int_0^{\infty}J(u-\mu)\rho_+(\mu)d\mu+
\int_0^{\infty}J(u+\mu+2B)\rho_+(\mu)d\mu
\label{7.1}
\end{equation}

\noindent
where $\rho_{j_0}$ is an unperturbed density describing the distribution of
the ground state strings (see table 4), $\hat{J}(w)=\frac{1}
{\hat{T}_{j_0,j_0}(u)+1}-1$ and $\rho_+(u)$ is density of holes induced by the
external magnetic field, and which are located outside of the completely
filled Fermi band.

In the eqn. (\ref{7.1}), the last term appearing in the right hand side can be
treated as a perturbation, provided $B\gg 1$. Iterating (\ref{7.1}), it is
straightforward to obtain

\begin{equation}
\left\{ \begin{array}{rll}
\rho_+(u) & = & \rho_+^{(0)}(u)+\rho_+^{(1)}(u)+\dots \nonumber \\
\rho_{j_0}(u+B) & = & \rho_+^{(0)}(u)+\int_0^{\infty}
J(u-\mu)\rho_+^{(0)}(0)(\mu)d\mu \nonumber \\
-\int_0^{\infty}J(u+\mu+2B)\rho_+^{(0)}(\mu)d\mu & = & \rho_+^{(1)}(u)+
\int_0^{\infty}J(u-\mu)\rho_+^{(1)}(\mu)d\mu \;,\;\;etc.
\end{array} \right.
\label{7.2}
\end{equation}

\noindent
and, for the magnetization $M$,

\begin{equation}
M=s- \sharp(Bethe\;\;roots)=M_0-2(1+\hat{J}(0))(\hat{\rho}_+^{(0)}(0)+
\hat{\rho}_+^{(1)}(0)+\dots)
\label{7.3}
\end{equation}

\noindent
where, as usual, "hat" designates Fourier transform and
$M_0=s-\hat{\rho}_{j_0}(0)$, which represents the nonvanishing ground state
magnetization ($h=0$), was given in ref. \cite{21}. \newline
One can recognize (\ref{7.2}) as Wiener-Hopf eqns. which can be solved in
standard fashion to give

\begin{eqnarray}
\hat{\rho}_+^{(0)}(w) & = & \sum_n\frac{\tilde{a}_n}{w-iw_n}\hat{J}_-(iw_n)
\frac{e^{-Bw_n}}{\hat{J}_+(w)} \nonumber \\
\hat{\rho}_+^{(1)}(w) & = & \sum_{n,j}\frac{\tilde{a}_n\tilde{b}_j}{w_j+w_n}
e^{-B(w_n+2w_j)}
\frac{\hat{J}_{-}(iw_n)\hat{J}_{-}(iw_j)}{w_j\hat{J}_+(0)}+d.p.
\label{7.4} \\
0 & < & w_{n(j)=1}<w_{n(j)=2}<\dots \nonumber
\end{eqnarray}

\noindent
Here, $\tilde{a}_n=res_{w=iw_n}\hat{\rho}_{j_0}(w)$;
$\tilde{b}_j=res_{w=iw_j}\hat{J}_+(w)$ and $d.p.$ stands for the double-poles
contribution which is not important for the following.
Kernels $\hat{J}_{\pm}(w)$ are defined as

\begin{eqnarray}
\hat{J}_+(w)=\hat{J}_-^{-1}(w)\;;\;\;\;\;
1+\hat{J}(w)=\frac{\hat{J}_+(w)}{\hat{J}_-(w)}\;;\;\;\;\;
\hat{J}_{\pm}(\infty)=1
\label{7.5}
\end{eqnarray}

\noindent
Explicit formulas for $\hat{J}_+$ are given below:

\begin{equation}
\left\{ \begin{array}{ccll}
\hat{J}_+(w) & = & \sqrt{\frac{p_0-1}{2\pi p_0}}
e^{i\frac{w}{\pi}(p_0\ln p_0-(p_0-1)\ln(p_0-1))}\times
\frac{\Gamma(\frac{iw}{\pi}
(p_0-1))\Gamma(\frac{1}{2}+i\frac{w}{\pi})}{\Gamma(\frac{iw}{\pi}p_0)}
& ;\;\;\;\;N\;even
\label{7.6} \\
\hat{J}_+(w) & = & \sqrt{\frac{p_0-\frac{1}{2}}{2\pi p_0}}
e^{i\frac{w}{\pi}(p_0\ln p_0-(p_0-\frac{1}{2})\ln(p_0-\frac{1}{2})-\ln\sqrt{2})
}\times\frac{\Gamma(\frac{iw}{\pi}(p_0-\frac{1}{2}))\Gamma(\frac{1}{2}+
i\frac{w}{\pi})}{\Gamma(\frac{iw}{\pi}p_0) \Gamma(\frac{1}{2}+i\frac{w}{2\pi})}
& ;\;\;\;\; N\;odd\;\;(N>3)
\end{array} \right.
\end{equation}

\noindent
Note that $\hat{J}_{\pm}(w)$ has singularities only in the upper (lower)-half
plane and is analytic and non-zero in the lower (upper)-half plane.

Before we move on, two comments are in order. \\
First, the Fourier image $\hat{\rho}_{j_0}(w)$ has two kinds of poles:

a) those, which are present for $s=j$ (regular spin) and

b) additional ones, which appear only when $s\neq j$ (i.e. in general
nilpotent case).\newline
It is precisely through the "b"-poles that nilpotency manifests itself in the
magnetic characteristic. As will be demonstrated shortly, the "b"-contribution
is $not\;invariant$ under the change of sign of the external magnetic
field. \\
Second, this comment concerns the pole structure of $\hat{J}_{\pm}(w)$-kernels.
The reader can easily convince oneself that this structure depends on whether
$\nu$ is even or odd.
Even though the difference in the analytical properties is not that important
for small $h$, it may become the relevant factor, once the strong magnetic
field is imposed. Note in this regard, that the Hilbert space of the
excitations (for $N$ even) is, indeed, sensitive to the $\nu$-parity, as was
shown in section 4.

To proceed further, one must establish the relation between Fermi momentum
$B$ and the magnetic field $h$. The easiest way to do this is to impose a
condition that the energy of a hole, located at the Fermi surface, vanishes,
i.e.
\begin{equation}
\epsilon_+(u=0)=0
\label{7.7}
\end{equation}

\noindent
where $\epsilon_+(u)$ is the energy of a hole, subject to

\begin{equation}
-4p_0I\rho_{j_0}(u+B)+\tilde{z}h=\epsilon_+(u)+\int_0^{\infty}
J(u-\mu)\epsilon_+(\mu)d\mu+\int_0^{\infty}J(u+\mu+2B)\epsilon_+(\mu)d\mu
\label{7.8}
\end{equation}

\noindent
with $\tilde{z}=p_0(\frac{p_0}{2})$ for $N$ odd (even).\\
Eqn. (\ref{7.8}) above is very similar to the eqn (\ref{7.1}) and can be
iterated in the same fashion to yield

\begin{equation}
\hat{\epsilon}_+(w)=-4p_0I(\hat{\rho}_+^{(0)}(w)+\hat{\rho}_+^{(1)}(w)+\dots)
+\frac{\tilde{z}h}{i(w-i\epsilon)}\frac{\hat{J}_-(0)}{\hat{J}_+(w)}
+\sum_j\frac{\tilde{b}_j\tilde{z}H\hat{J}_-(0)\hat{J}_-(iw_j)}
{iw_j\hat{J}_+(w)i(w-iw_j)}e^{-2Bw_j}+\dots
\label{7.9}
\end{equation}

\noindent
In terms of $\hat{\epsilon}_+(w)$ one can rewrite (\ref{7.7}) as follows:

\begin{equation}
\lim_{w\rightarrow\infty}iw\hat{\epsilon}_+(w)=0
\label{7.10}
\end{equation}
\newline
It is reasonable to expect that the effect of the interplay between "internal"
and external magnetic fields will be visible in the lowest order of the
perturbation theory, if the number of "affected" degrees of freedom is
comparable with the total number of degrees of freedom, i.e. if $p_0$ is not
too large. To show that this is, indeed, the case let us calculate
magnetization $M$ for $p_0=5$. \newline
Making use of (\ref{7.3}, \ref{7.4}, \ref{7.10}) one finds

\begin{equation}
M(h)=M_0+\frac{h}{2\pi I}+(\frac{h}{I})^2\frac{2}{45\pi^2\cos(\pi\delta s)}
\{\frac{\Gamma^2(\frac{1}{4})}{\Gamma(\frac{3}{4})\Gamma(-\frac{5}{4})}+
\frac{32\sin(\pi\delta s)}{\pi}\}\dots
\label{7.11}
\end{equation}

\noindent
where $\delta s=s-j$.

In the discussion above it was assumed that $h>0$. The best way to treat $h<0$
case is to exploit "charge conjugation" authomorphic properties of our
nilpotent hamiltonian $H$, namely
\begin{eqnarray}
H(\delta s,-h) & \rightarrow & H(-\delta s,h) \\
\label{7.12}
s & \rightarrow & N'-1-s \nonumber
\end{eqnarray}

\noindent
Note that these properties are the direct consequences of (\ref{3.37}).\\
Now it is obvious that
\begin{equation}
M(-|h|)=M_0+\frac{-|h|}{2\pi I}+(\frac{|h|}{I})^2\frac{2}
{45\pi^2\cos(\pi\delta s)}
\{\frac{\Gamma^2(\frac{1}{4})}{\Gamma(\frac{3}{4})\Gamma(-\frac{5}{4})}-
\frac{32\sin(\pi\delta s)}{\pi}\}\dots
\label{7.13}
\end{equation}
\newline
\noindent
Comparing eqns. (\ref{7.11}) and (\ref{7.13}), one cannot help noticing that
the coefficient
of the quadratic term does depend on the sign of the magnetic field. Clearly,
this effect bears the signature of the irregular irrep., since it only exists
when the systems is away from the commensurable point. For $p_0\gg 1$ one
should keep sufficiently many terms in the perturbation expansion to "catch"
this highly nontrivial effect.

Finally, let us present $M(h)$ for the generic $p_0\gg 1$

\begin{equation}
M-M_0=\frac{\tilde{z}h}{2p_0Iw_{n=1}}\{1+(\frac{\tilde{z}\hat{J}_-(0)h}
{4ip_0I\hat{J}_-(iw_{n=1})\tilde{a}_{n=1}})^{\frac{2w_{j=1}}{w_{n=1}}}g\}
+\dots\
\label{7.14}
\end{equation}

\noindent
where $g=\frac{2w_{n=1}\hat{J}_-(iw_{j=1})\tilde{b}_{j=1}}
{iw_{j=1}(w_{j=1}+w_{n=1})}$.\newline
It is interesting to observe that the first term in the expansion (\ref{7.14})
is always universal (which is consistent with universality of the central
charge) and that next to leading term depends on $\delta s$. The presence of
the fractional powers of $h$ in the formula above should be taken as an
indication that $M$ has nontrivial analytical properties.

To conclude this section, we would like to comment that the usual criteria of
the small magnetic field $h\ll I$ does not quite work in our case. Indeed,
note that $h$ in the second term of (\ref{7.14}) appears in combination
$\frac{h}{I\tilde{a}_{n=1}(\delta s)}$. Keeping $h$ fixed and varying $s$
towards the
orbifold point can make this combination very large. In fact, it can
be shown that for arbitrary small $\frac{h}{I}$, there is a region
adjacent to the orbifold point where the system is in completely ordered
ferromagnetic state. We hope to report on the strong magnetic field properties
of our system in the future publications.

\chapter*{  Concluding Remarks}

In this paper we have analysed the special features of integrable
higher spin $XXZ$ chains resulting from imposing spin-anisotropy
commensurability. This study can be considered from different points of view
each one providing a direction of research we feel worth pursuing. We sketch
some of them currently under study:

i) Generalization of the concept of free fermions and $N=2$ SUSY.
Starting with
the work of Onsager free fermions are the basic tool to solve a large
class of 2D statistical models. The simplest generalization is
to work with $Gl(1,1)_{\hat{q}}$-invariant chains or equivalently with
nilpotent
deformations of $SU(2)_q$-chains with $q^4 =1$. We found
for commensurable chains with $q^N =1$ ($N$ even)
that the spectrum of excitations consists of free fermions, "dressed" by the
RSOS degrees of freedom,
with their "vertex dynamics" completely determined by the
$SU(2)_q\;(q^4=1)$ quantum symmetry. This symmetry underlies a new non-local
realization of $N=2$ SUSY. The connection between free fermions and
$N=2$ models may shed some new light on the Painlev\'e\`e structure
recently found in the study of $N=2$ models \cite{37}.
Our results for the $q^N =1$ ($N$ odd) chains show the way to find a
definition of free parafermions. Some hints of the non trivial structure of
such systems already appear in the observed failure of naive bootstrap.

ii) Integrability in higher dimensions. $N=2$ SUSY theories,
free fermion models and their generalizations are natural
candidates for a direct 3-dimensional interpretation.
A preliminary research in this direction was started in \cite{38}.
The integrable deformations associated with nilpotent irreps
we have considered in this paper allow, in some simple cases,
the interpretation in terms of dimensionally reduced 3D-models.

iii) Selective magnetism. Our models can be interpreted in terms of an
"internal" magnetic field which selectively affects available degrees of
freedom. As a consequence, the response of these systems to the external
magnetic field exhibits asymmetry under the change of sign of the
external field.
Although, this effect is relatively small, it can be taken as a signature of
an extra dimension in the spirit of comment (ii).

We believe that the study of the commensurable spin chains provides good
opportunities to link a variety of topics which are ripe for unification.

\appendix
\chapter{Hermiticity Regions for Nilpotent Hamiltonians}

\setcounter{equation}{0}

In this appendix we shall describe the intervals of the generalized
spin $s$ for which the corresponding hamiltonian with anisotropy
$\gamma = 2 \pi M/N$ is hermitian. We shall assume that $N$ and $M$ are
relative primes such that $ 1 \leq M < N$. Then the problem is
to find the regions where the variable $x = 2s$
($ 0 \leq x \leq N$) satisfies

\begin{equation}
v_s \;\;\; sin 2 \pi \frac{M}{N} (r+1) \;\; \;sin 2 \pi \frac{M}{N} (x -r)
> 0
\label{h1}
\end{equation}

\noindent
for $ r = 0, 1 , \dots, N'-2$.

Considering a few examples, one extracts the following facts:

i) There are $M$ intervals with positive parity and $M$ intervals
with negative parity which alternate.

ii) The width of each interval ($x_1,x_2$) is given by:

\begin{equation}
x_2 - x_1= \left\{
\begin{array}{cl}
1/M & N\;odd \\
2/M & N\;even
\end{array}
\right.
\label{h2}
\end{equation}

iii) The orbifold points $ \lambda_1 = q^{x_1}$ and $ \lambda_2
= q^{x_2}$  that delimitate the intervals are related by charge
conjugation:

\begin{equation}
\lambda_2 = (\lambda_1)_C = q^{-2} \lambda_1^{-1}
\label{h3}
\end{equation}

{}From these data we want to find the values of $x_1$ and $x_2$
which, in turn, imply the patterns of breaking of the nilpotent
irreps at the orbifold points. From (\ref{h2}) and (\ref{h3}) we get:

\begin{eqnarray}
N\;odd: & k N -(1 + x_1)  2 M & = 1 \nonumber \\
N\;even: & \frac{k}{2} N - (1 + x_1 ) M & =1
\label{h4}
\end{eqnarray}

\noindent
where k is some integer. These equations admit several solutions
for $x_1$ which differ by $ \Delta x_1 = \frac{N}{2 M}$.
We shall choose the solution for which $d = 1 + x_1$ gives us
the dimension of one of the irreps at the orbifold point. The
other is equal to $ N' - d$. Both cases in (\ref{h4}) can be
treated simultaneously if one recalls the definition of $N'$
and one defines $M'$ as:

\begin{eqnarray}
N\;odd: & N' = N, & M' = 2M' \nonumber \\
N\;even: & N' = \frac{N}{2}, & M= M'
\label{h5}
\end{eqnarray}

\noindent
Using (\ref{h5}), equation (\ref{h4}) becomes:

\begin{equation}
k \; N' - d \; M' = 1
\label{h6}
\end{equation}

Since $N'$ and $M'$ are relative primes, this equation
has a unique solution for $k$ and $d$ (both integers) which
can be found using the fraction expansion of $p_0 \equiv
\frac{N'}{M'} = \frac{\pi}{\gamma}$:

\begin{equation}
p_0 = \frac{N'}{M'}= \nu_1 +
\frac{1}{\cdots + \frac{1}{\nu_{\alpha}}}
\label{h7}
\end{equation}

The integers $\nu_i  (i= 1, \dots, \alpha)$ can be computed using
the Euclid algorithm. In terms of them one  defines the reduced
fractions:

\begin{equation}
\delta_1 = \nu_1, \delta_2= \nu_1 + \frac{1}{\nu_2},
\delta_3 = \nu_1 + \frac{1}{\nu_2 + \frac{1}{\nu_3}},\dots,
\delta_{\alpha}= \frac{N'}{M'}
\label{h8}
\end{equation}

\noindent
which can also be written as

\begin{equation}
\delta_i = \frac{y_i}{z_i}
\label{h9}
\end{equation}

\noindent
where $y_i$ and $z_i$ are computed from the recursion formulas:

\begin{eqnarray}
y_0 =1, & y_1 = \nu_1,& y_i = \nu_i y_{i-1} + y_{i-2} \nonumber \\
z_0 =0, & z_1 = \nu_1,& z_i = \nu_i z_{i-1} + z_{i-2}
\label{h10}
\end{eqnarray}

\noindent
Notice that $y_{\alpha}= N'$ and $ z_{\alpha} = M'$.
One can easily prove by induction:

\begin{equation}
y_i z_{i-1} - z_i y_{i-1} = (-1)^i \;, i \geq 1
\label{h11}
\end{equation}

Setting $i= \alpha$ in (\ref{h11}) we get:

\begin{equation}
N' \; z_{\alpha -1} - M'\; y_{\alpha -1} = (-1)^{\alpha}
\label{h12}
\end{equation}

\noindent
and comparing with (\ref{h6}) we, finally, obtain:

\begin{eqnarray}
\alpha \;\;even: & k = z_{\alpha -1} ,& d= y_{\alpha -1} \nonumber \\
\alpha \;\;odd: & k = M' - z_{\alpha -1} ,& d=N'-  \;\;y_{\alpha -1}
\label{h13}
\end{eqnarray}

Therefore, for $\alpha $ even or odd, the dimensions of the regular
irreps into which the nilpotent irrep breaks are $y_{\alpha}$ and
$N' - y_{\alpha}$. Furthermore, we shall now show that both
numbers $y_{\alpha} $ and $N' - y_{\alpha}$ are Takahashi numbers.
Indeed, the Takahashi numbers are the pairs ($n_j,v_j$)
which give the string length and parity of the allowed strings
for a given anisotropy $\gamma$ and can be computed using the formula
\cite{17}:
\begin{eqnarray}
n_j = y_{i-1} + (j-m_i)y_i & for &m_i \leq j \leq m_{i+1}
\label{h14}
\end{eqnarray}

\noindent
where $ m_0 = 0 ,m_i = \sum^i_{k=1} \nu_k$ and the parities $v_i$ are:

\begin{equation}
v_1= 1, \; v_{m_1}=-1, \; v_j= exp \left(i \pi \left[ \frac{n_j
-1}{p_0} \right] \right) \; for \;\;j \neq1,m_1
\label{h15}
\end{equation}

Here $[x]$ means the integer part of $x$.
For rational values of $p_0$ the index $i$ in (\ref{h14}) runs
from 0 to $\alpha$ and one has an an extra condition besides (\ref{h14}):

\begin{eqnarray}
n_j < \ell && \forall j
\label{h16}
\end{eqnarray}

The allowed values of $n_j$ given by (\ref{h14}) form the $i$-th Takahashi
zone. Thus, there are $\alpha +1$ zones. Condition (\ref{h16})
implies that the last zone, i.e. $i=\alpha$, has only one string
of length $n_{m_{\alpha}}= y_{\alpha-1}$. On the other hand,
using (\ref{h10}) one gets for the last string of the zone $i=\alpha-1$
the length $n_{m_{\alpha}-1}= N' - y_{\alpha -1}. $

We have proved that the lengths of the two last strings
in the Takahashi-Suzuki labeling coincide with the dimensions of
the regular irreps into which the nilpotent irrep breaks at the
orbifold points. This situation is quite different from the one considered
by Kirillov and Reshetikhin \cite{9} (see also ref. \cite{24}) in their study
of higher spin $XXZ$ models. They study situations where the anisotropy
$\gamma$ is such that $2j +1$ is a Takahashi number, where $j$ stands for the
spin of the chain. We observe that in our case $2j +1 = N'$ is not
a Takahashi number, although, the irreducible pieces into which it
breaks at the orbifold points are, indeed, the Takahashi numbers.

Below we show some patterns of breaking for low values of $N$.

\begin{center}
\begin{tabular}{|c|c|l|}
\hline
$N$ & $M$ & Breaking \\
\hline
4 & 1 & 2 $\rightarrow$ 1 $\oplus$ 1 \\
\hline
3 & 1, 2 & 3 $\rightarrow$ 2 $\oplus$ 1 \\
6 & 1, 5 & 3 $\rightarrow$ 2 $\oplus$ 1 \\
\hline
8 & 1,3,5,7 & 4 $\rightarrow$ 3 $\oplus$ 1 \\
\hline
5 & 1,4 & 5 $\rightarrow$ 3 $\oplus$ 2 \\
5 & 2,3 & 5 $\rightarrow$ 4 $\oplus$ 1 \\
10 & 1,9 & 5 $\rightarrow$ 4 $\oplus$ 1 \\
10 & 3,7 & 5 $\rightarrow$ 3 $\oplus$ 2 \\
\hline
12 & 1,11 & 6 $\rightarrow$ 5 $\oplus$ 1 \\
12 & 5,7  & 6 $\rightarrow$ 5 $\oplus$ 1 \\
\hline
\end{tabular}
\begin{center}
Table 5.
\end{center}
\end{center}

\chapter{The Inversion Formulas for Kernels $T_{j,k}$}

Starting with the formulas (\ref{4.8}), it is straightforward (though tedious)
to verify the following inversion relations exhibited in the table below.

\begin{center}
\begin{tabular}{|c|c|}
\hline
$N$ even & $N$ odd \\
\hline
$\hat{A}_{j,k}=\frac{1}{2chw} \{ \hat{A}_{j+1,k}+\hat{A}_{j-1,k} \}+
\delta_{j,k}-\frac{\delta_{j,\nu-2}\delta_{k,\nu}}{2chw}\;;$
& $\hat{A}_{j,k}=\frac{1}{2chw}\{ \hat{A}_{j+1,k}+\hat{A}_{j-1,k} \}+
\delta_{j,k} \;;\;\;$ $1\leq j\leq\nu-2$ \\
$1\leq j\leq\nu-2$.
& $(1-\frac{1}{2chw})\hat{A}_{\nu-1,k}=\frac{1}{2chw}\hat{A}_{\nu-2,k}+
\frac{1}{2ch\frac{w}{2}}\hat{A}_{\nu,k}+\delta_{\nu-1,k}$ \\
$\hat{A}_{\nu-1,k}=\frac{1}{2chw}\hat{A}_{\nu-2,k}+\delta_{\nu-1,k}$
& $-\hat{A}_{\nu,k}=\frac{1}{2ch\frac{w}{2}}\{ \hat{A}_{\nu-1,k}-
\hat{A}_{\nu+1,k} \} + \delta_{\nu,k}-
\frac{\delta_{j,\nu}\delta_{k,\nu+2}}{2ch\frac{w}{2}}$ \\
$\hat{A}_{\nu,k}=\frac{1}{2chw}\hat{A}_{\nu-2,k}+\delta_{\nu,k}$
& $\hat{A}_{\nu+1,k}=\frac{1}{2ch\frac{w}{2}}\hat{A}_{\nu,k}-
\delta_{\nu+1,k}$ \\
& $\hat{A}_{\nu+2,k}=-\frac{1}{2ch\frac{w}{2}}\hat{A}_{\nu,k}+\delta_{\nu+2,k}$
\\
\hline
\end{tabular}
\begin{center}
Table 6.
\end{center}
\end{center}

\noindent
where $\nu=\frac{N}{2}$ ($\frac{N-1}{2}$) for $N$ even ($N$ odd); symbol "hat"
above kernel $A$ designates Fourier transform and, finally, the connection
between kernels $A$ and $T$ is given by:

$$
A_{j,k}(u)=T_{j,k}(u)+(-1)^{r_j}\delta_{j,k}\delta(u) \nonumber \\
$$

The inversion formulas presented in the table above are quite similar to those
discussed in ref. \cite{17}. The main difference is related to the fact that
ref. \cite{17}
deals with the generic anisotropy $\gamma$ and as a result infinitely many
Takahashi zones enter into the analysis. Since in our case, $q$ is a root of
unity, the number of relevant Takahashi zones is always finite. This fact is
reflected in the presence of extra $\delta$ function pieces which appear in our
inversion relations.

\chapter*{Figure Captions}

\begin{tabular}{lcl}
Fig. 1. & & Hermiticity Regions \\
\\
Fig. 2. & & Spin-Anisotropy Plane and the Commensurable Point \\
\\
Fig. 3. & & Bratteli Diagrams for $A_{2j+2}$ graph and the Paths for $p_0$ even
(odd) \\
\\
Fig. 4. & & $(p=2, r=6)$ RSOS graph \\
\end{tabular}

\end{document}